\documentclass[aip, jcp, 12pt]{revtex4-1}
\usepackage{amsmath}
\usepackage{amssymb}
\usepackage{bm}
\usepackage{color}
\usepackage{graphicx}
\usepackage{hyperref}
\usepackage{natbib}
\usepackage{scalerel}
\usepackage{siunitx}

\newcommand{\D}{\operatorname{d\!}}
\newcommand{\E}{\operatorname{e}}
\newcommand{\bra}[1]{\langle #1 |}
\newcommand{\braket}[2]{\langle #1 | #2\rangle}
\newcommand{\ket}[1]{| #1 \rangle}
\newcommand{\normord}[1]{\mathopen: #1 \mathclose:}
\newcommand{\antisymm}{\mathop{\scalerel*{\mathcal{A}}{\textstyle\sum}}\displaylimits}

\begin{document}
\title{Addition and removal energies of circular quantum dots}

\author{Fei Yuan}
\email{yuan@nscl.msu.edu}
\homepage{https://people.nscl.msu.edu/~yuan}
\affiliation{National Superconducting Cyclotron Laboratory and Department of Physics and Astronomy, Michigan State University, East Lansing, MI 48824, USA}
\author{Samuel J. Novario}
\affiliation{National Superconducting Cyclotron Laboratory and Department of Physics and Astronomy, Michigan State University, East Lansing, MI 48824, USA}
\author{Nathan M. Parzuchowski}
\affiliation{National Superconducting Cyclotron Laboratory and Department of Physics and Astronomy, Michigan State University, East Lansing, MI 48824, USA}
\affiliation{The Ohio State University, Columbus, OH 43210, USA}
\author{Sarah Reimann}
\affiliation{Department of Chemistry, Hylleraas Centre for Quantum Molecular Sciences, University of Oslo, N-0316 Oslo, Norway}
\author{S. K. Bogner}
\affiliation{National Superconducting Cyclotron Laboratory and Department of Physics and Astronomy, Michigan State University, East Lansing, MI 48824, USA}
\author{Morten Hjorth-Jensen}
\affiliation{National Superconducting Cyclotron Laboratory and Department of Physics and Astronomy, Michigan State University, East Lansing, MI 48824, USA}
\affiliation{Department of Physics, University of Oslo, N-0316 Oslo, Norway}

\begin{abstract}
  We present and compare several many-body methods as applied to two-dimensional quantum dots with circular symmetry.  We calculate the approximate ground state energy using a harmonic oscillator basis optimized by Hartree--Fock (HF) theory and further improve the ground state energy using two post-HF methods: in-medium similarity renormalization group (IM-SRG) and coupled cluster with singles and doubles (CCSD).  With the application of quasidegenerate perturbation theory (QDPT) or the equations-of-motion (EOM) method to the results of the previous two methods, we obtain addition and removal energies as well.  Our results are benchmarked against full configuration interaction (FCI) and diffusion Monte Carlo (DMC) where available.  We examine the rate of convergence and perform extrapolations to the infinite basis limit using a power-law model.
\end{abstract}

\maketitle

\section{Introduction}

The behavior of strongly confined electrons is a major problem for many-body theory.  The prototypical system is that of quantum dots, also known as ``artificial atoms'', in which electrons are confined within artificially constructed semiconducting heterostructures.  Such nanoscale systems are highly relevant as both theoretical models and experimental observations can readily probe quantum phenomena such as tunneling, entanglement, magnetization, and symmetry breaking \cite{reimann2002,engel1993,BIRMAN20131}.  Moreover, unlike many physical systems, quantum dots enjoy the benefit of being highly tunable through changes in the external field or the structure of the confining material.  This allows experiments to quantify the impact of quantum effects at different levels of correlation.

The ground states of quantum dots exhibit shell structures and magic numbers similar to those of atoms and nuclei \cite{tarucha1996}.  Thus, they provide a rare opportunity to study electronic systems without the influence of atomic nuclei.  They also provide a testbed for the development of many-body methods for other systems with similar geometry, such as nuclei or neutron drops \cite{PhysRevC.84.044306}.

Beyond their theoretical relevance, there are numerous potential applications of quantum dots.  The electrical and optical properties of quantum dots are particularly useful for solar cells \cite{jenks:013111,doi:10.1021/cr900289f} and laser technology \cite{strauf2010,5075760}.  They have potential medical applications in diagnosis and treatment \cite{Ben-Ari02042003}.  And of course, quantum dots are also promising candidates for the physical realization of quantum computing \cite{PhysRevA.57.120}.

In general, quantum dot systems are not analytically solvable, with the exception of two-electron systems or systems with specific strengths of the external field \cite{PhysRevA.48.3561,10.1088/0305-4470/27/3/040}.  Hence, in practical applications one can only hope for a numerical solution through some \textit{many-body method} or a combination thereof.

One of the most accurate techniques for solving many-body systems is that of exact diagonalization, also known as the full configuration interaction method.  Unlike most methods, it is assured to converge to the exact answer as the size of the finite basis is increased to infinity (the so-called \textit{infinite-basis limit}).  Despite this significant advantage, it is often not possible to perform exact diagonalization due to its factorially increasing cost with respect to the basis size and the number of particles.  This led to the development of more cost-effective methods that trade varying amounts of accuracy for varying amounts of speed.

In this paper, we analyze the application of several methods to the quantum dot systems, including the Hartree--Fock (HF) method \cite{hartree_1928,Fock1930}, M\o ller--Plesset (MP) perturbation theory \cite{1934PhRv...46..618M}, the in-medium similarity renormalization group (IM-SRG) method \cite{Hergert2016165}, coupled cluster (CC) theory \cite{PhysRevB.67.045320,heidari:114708,PhysRevB.84.115302}, quasidegenerate perturbation theory (QDPT) \cite{0022-3700-7-18-010,Kvasnicka1974}, and equations-of-motion (EOM) methods \cite{RevModPhys.40.153,StantonBartlettEOM,EMRICH1981379}.  We compare them against variational and diffusion Monte Carlo (VMC and DMC) \cite{PhysRevB.68.035304,PhysRevB.62.8120,PhysRevB.84.115302,PhysRevB.54.4780} and full configuration interaction (FCI) theory \cite{olsen2013thesis,JJAP.36.3924,PhysRevB.56.6428,2008arXiv0810.2644K,rontani:124102} results, where available, from existing literature.

Of these methods, IM-SRG is a recently developed technique that has shown significant promises.  Similarity renormalization group (SRG) methods \cite{PhysRevD.48.5863,PhysRevD.49.4214} are a family of methods that transform the Hamiltonian into a band- or block-diagonal form through a continuous sequence of unitary transformations.  The goal of such a transformation is to reduce the coupling between a subspace of interest -- such as the ground state or a set of low-lying states -- and the remaining Hilbert space.  It has successfully been applied to systems with various underlying potentials to calculate their binding energy and other observables, especially in nuclear theory \cite{ScottSRG,PhysRevC.75.061001,SRGThreeDim}.  The \emph{in-medium} SRG (IM-SRG) method adapts the SRG approach to evolve Hamiltonians in a truncated Fock space that is centered around an approximate reference state rather than the physical vacuum, which substantially reduces the importance of the computationally costly higher-body operators.

This article is organized as follows: Section \ref{sec:formalism} introduces first (\ref{subsec:modelHamiltonian}) the Hamiltonian and basis we use to model circular quantum dots, and gives afterwards an overview of the \textit{ab initio} many-body methods used in this paper: HF, IM-SRG, CCSD, QDPT, EOM.  Our results are presented in Section \ref{sec:results}.  We analyze and compare the differences between the various methods and also between different quantum dot systems.  We examine the utility of extrapolation techniques in improving the precision of our results.  Section \ref{sec:conclusions} concludes our work and gives perspectives for future work.

\section{Formalism}
\label{sec:formalism}

\subsection{The model Hamiltonian}
\label{subsec:modelHamiltonian}

We shall model the circular quantum dot system as a collection of $N$ nonrelativistic electrons of mass $m$ in two-dimensional space, trapped by an external harmonic-oscillator potential of the form $m \omega^2r^2 / 2$, where $\omega$ is its angular frequency and $r$ is the radial distance from the center.  The electrons interact with each other through the standard Coulomb interaction $e^2 / (4 \pi \epsilon R)$, where $e$ is the electron charge, $\epsilon$ is the permittivity of the medium, and $R$ is the distance between the two interacting electrons.

For simplicity, we will use atomic units, choosen such that $\hbar \equiv m \equiv e \equiv 4 \pi \epsilon \equiv 1$, reducing the parameters of the system from $(N, m, \omega, e, \epsilon)$ to just $(N, \omega)$.  Hence, all energies and frequencies are presented in hartrees and hartrees per $\hbar$ respectively.

In these units, the many-body problem is described by the Hamiltonian
\begin{align} \label{eq:fullhamiltonian}
  \hat H &\equiv \hat{H}_1 + \hat{H}_2 & \hat{H}_1 &\equiv \sum_{\alpha = 1}^N \hat{h}_\alpha & \hat{H}_2 &\equiv \sum_{\alpha = 1}^N \sum_{\beta = 1}^{\alpha - 1} \frac{1}{\hat{\bm r}_\alpha - \hat{\bm r}_\beta},
\end{align}
where $\hat{H}_1$ and $\hat{H}_2$ are its one- and two-body parts respectively and, for the $\alpha$-th particle, $\hat{\bm r}_\alpha$ is its position operator and  $\hat{h}_\alpha$ is its single-particle harmonic-oscillator Hamiltonian,
\begin{align*}
  \hat{h} \equiv -\frac{1}{2} \hat{\bm{\nabla}}^2 + \frac{1}{2} \omega^2 \hat{\bm{r}}^2.
\end{align*}

The noninteracting Hamiltonian $\hat{H}_1$ can be solved as $N$ independent single-particle problems involving $\hat{h}$ with easy analytic solutions in Cartesian form.  However, to exploit circular symmetry, we use instead the Fock--Darwin states $F_{n m_\ell}$, which favor polar coordinates $\bm{r} = (\rho, \varphi)$ and conserve orbital angular momentum $\hat{L}_{\mathrm{z}} \equiv -\mathrm{i} \frac{\partial}{\partial \varphi}$.  They are defined as\cite{lohne2010coupled}
\begin{align}
  F_{n m_\ell}(\rho, \varphi) &\equiv \sqrt\omega R_{n |m_\ell|}(\sqrt \omega \rho) \times \frac{1}{\sqrt{2 \pi}} \mathrm{e}^{\mathrm{i} m_\ell \varphi} \label{eq:fockdarwin}, \\
  R_{n m}(\varrho) &\equiv \sqrt{\frac{2 \times n!}{(n + m)!}} \mathrm{e}^{-\varrho^2 / 2} \varrho^m L_n^{(m)}(\varrho^2), \notag
\end{align}
where $L_n^{(\alpha)}$ denotes the generalized Laguerre polynomial \cite{NIST:DLMF} of degree $n$ and parameter $\alpha$,
\begin{align*}
  L_n^{(\alpha)}(u) \equiv \frac{1}{n!} u^{-\alpha} \mathrm{e}^u \frac{\mathrm{d}^n}{\mathrm{d} u^n} (\mathrm{e}^{-u} u^{\alpha + n}).
\end{align*}
The states are distinguished by two quantum numbers: the principal quantum number $n$, a nonnegative integer related to the degree of the Laguerre polynomial, and the orbital angular momentum projection $m_\ell$, the integer eigenvalue of $\hat{L}_{\mathrm{z}}$.

\begin{figure}
  \includegraphics{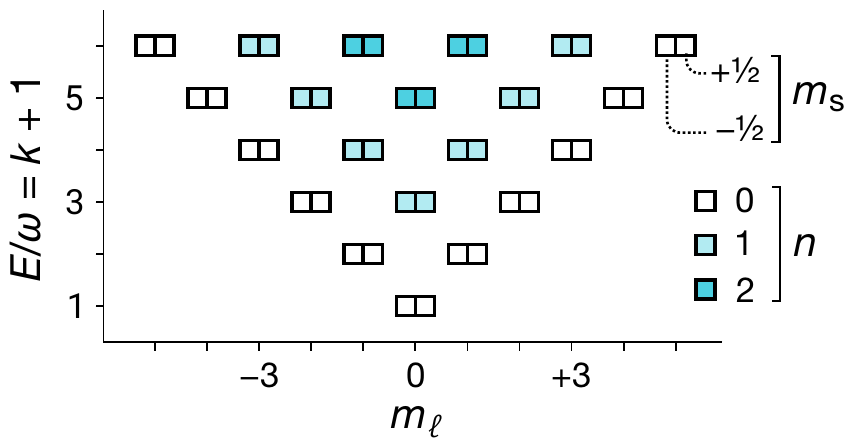}
  \caption{(Color online) The 42 lowest single-particle states (the first 5 shells) in the 2D harmonic oscillator basis.  Each box represents a single-particle state arranged by $m_\ell$, $m_{\mathrm{s}}$, and energy, and the up/down arrows indicate the spin of the states.  Within each column, the principal quantum number $n$ increases as one traverses upward.}
  \label{fig:shell-structure}
\end{figure}

Single-particle states of spin-$\frac{1}{2}$ electrons also contain a spin component,
\begin{align} \label{eq:singleparticlestate}
  \langle \rho \varphi m_{\mathrm{s}}' |n m_\ell m_{\mathrm{s}}\rangle \equiv F_{n m_\ell}(\rho, \varphi) \delta_{m_{\mathrm{s}}^{} m_{\mathrm{s}}'},
\end{align}
where $m_{\mathrm{s}} = \pm\frac{1}{2}$ is the spin projection quantum number and $\delta$ is the Kronecker delta.

The energy of each single-particle state $|n m_\ell m_{\mathrm{s}}\rangle$ is given by
\begin{align} \label{eq:energysingleparticlestate}
  \varepsilon_{n m_\ell m_{\mathrm{s}}} \equiv (2 n + |m_\ell| + 1) \omega.
\end{align}
They are degenerate with respect to both spin projection $m_{\mathrm{s}}$ and \textit{shell index} $k$,
\begin{align} \label{eq:shell_index}
  k \equiv 2 n + |m_\ell|,
\end{align}
which labels each shell from zero.  The shell structure is illustrated in Fig.\ \ref{fig:shell-structure}.

Fermionic $N$-particle eigenstates of the one-body Hamiltonian $\hat{H}_1$ can be explicitly constructed as Slater determinants $|p_1 \ldots p_N\rangle$ of the single-particle states $|p_1\rangle, \ldots, |p_N\rangle$ from Eq.\ \eqref{eq:singleparticlestate}, where $|p\rangle$ is an abbreviation of $|n m_\ell m_{\mathrm{s}}\rangle$.  The Slater determinant $|p_1 \ldots p_N\rangle$ is said to \textit{occupy} the single-particle states $|p_1\rangle, \ldots, |p_N\rangle$.

When the number of particles $N$ satisfies $N = K_{\mathrm{F}} (K_{\mathrm{F}} + 1)$ for some nonnegative integer $K_{\mathrm{F}}$, there would be just enough particles to form a closed-shell Slater determinant, leading to a nondegenerate, well-isolated ground state.  The values of $N$ at which this occurs are often termed \textit{magic numbers}, and $K_{\mathrm{F}}$ is the \textit{number of filled shells} (or more abstractly the ``Fermi level'').  A single-particle state is occupied in the ground state Slater determinant if and only if $k < K_{\mathrm{F}}$, where $k$ is its shell index as defined in Eq.\ \eqref{eq:shell_index}.

In this paper, we use the letters $p, q, r, s, \ldots$ to label arbitrary \textit{single-particle states} in the basis.  In the discussion of many-body methods that rely a single, distinguished Slater determinant $\Phi$, known as the \textit{reference state}, we use the $i, j, k, l, \ldots$ as placeholders for the occupied states of $\Phi$, and $a, b, c, d, \ldots$ for the unoccupied states.  A summation such as $\sum_p$ is understood as a summation over every single-particle basis state $p$.  In contrast, $\sum_i$ only sums over occupied states and $\sum_a$ over unoccupied ones.

It is generally more convenient to describe many-body methods within the formalism of second quantization; see for example Ref.\ \cite{shavitt2009many} for details.  For fermions, this involves annihilation operators $\hat a_p$ and creation operators $\hat a_p^\dagger$ that, by construction, satisfy the canonical anticommutation relations.  The one- and two-body Hamiltonian operators $\hat{H}_1$ and $\hat{H}_2$ are rewritten as
\begin{align} \label{eq:second_quantized_hamiltonian}
  \hat{H}_1 &= \sum_{p q} \langle p | \hat{H}_1 | q \rangle \hat a_p^\dagger \hat a_q^{}, &
  \hat{H}_2 &= \frac{1}{4} \sum_{p q r s} \langle p q | \hat{H}_2 | r s \rangle \hat a_p^\dagger \hat a_q^\dagger \hat a_s^{} \hat a_r^{},
\end{align}
where the quantities $\langle p | \hat{H}_1 | q \rangle$ and $\langle p q | \hat{H}_2 | r s \rangle$ are referred to as \textit{matrix elements} and fully characterize the two operators in the basis.  Specifically, $\langle p q | \hat{H}_2 | r s \rangle$ are \emph{antisymmetrized} matrix elements, which should not to be confused with the (non-antisymmetric) interaction integral,
\begin{align} \label{eq:interactionintegral}
  v_{p_1 p_2 p_3 p_4} \equiv \int \langle p_1 | \bm{r} \rangle \langle p_2 | \bm{r}'\rangle V(\bm r, \bm r') \langle \bm{r} | p_3 \rangle \langle \bm{r}' | p_4 \rangle \D \bm{r} \D \bm{r}'.
\end{align}
where for the standard Coulomb interaction we would have $V(\bm r, \bm r') = 1 / |\bm r - \bm r'|$.  In practice, antisymmetrized matrix elements are often computed from interaction integrals by antisymmetrization, $\langle p q | \hat{H}_2 | r s \rangle = v_{p q r s} - v_{p q s r}$.

We now discuss the various techniques we use to solve the interacting many-body problem.  The principal difficulty stems from the fact that, due to the presence of interactions, the exact solution is in general \emph{not} a single Slater determinant built from the single-particle states.  However, for closed-shell systems, if we assume the interaction alters the behavior of the system only mildly, we can use a Slater determinant of the noninteracting system as a starting point and apply various methods to improve the accuracy of the solution.  The single-particle states of $\hat{H}_1$ serve as the \textit{single-particle basis} of the many-body methods that we shall describe.

\subsection{Hartree--Fock method}
\label{subsec:HartreeFockmethod}

One of the simplest corrections is that of the Hartree--Fock (HF) method, also known as the self-consistent field (SCF) method.  Using the variational principle, one can obtain an approximate ground state of a closed-shell system by minimizing the energy expectation value $E_{\Phi}$ with respect to some Slater determinant $|\Phi\rangle$ in an unknown single-particle basis.  We shall denote each state $|p'\rangle$ in this unknown basis by a primed label $p'$.  We assume each unknown state $|p'\rangle$ is built from a linear combination of known basis states $|p\rangle$ with an unknown unitary matrix of coefficients $C_{p p'}$.

The goal is to find the coefficients $C_{p p'}$ that minimize the \textit{Hartree--Fock energy} $E_{\Phi}$,
\begin{align}
  E_{\Phi} &= \sum_{i'} \langle i' | \hat{H}_1 | i' \rangle + \frac{1}{2} \sum_{i' j'} \langle i' j' | \hat{H}_2 | i' j' \rangle \label{eq:hfenergy},
\end{align}
where
\begin{align}
  \langle p' | \hat{H}_1 | q' \rangle &= \sum_{p q} C_{p p'}^* \langle p | \hat{H}_1 | q \rangle C_{q q'}^{}, \label{eq:hftransform1} \\
  \langle p' q' | \hat{H}_2 | r' s' \rangle &= \sum_{p q r s} C_{p p'}^* C_{q q'}^* \langle p q | \hat{H}_2 | r s \rangle C_{r r'}^{} C_{s s'}^{}, \label{eq:hftransform2}
\end{align}
Using the method of Lagrange multipliers, the minimization problem reduces to a nonlinear equation -- the self-consistent \textit{Hartree--Fock equations}:
\begin{align} \label{eq:hartreefock}
  \sum_q F_{p q} C_{q p'} = C_{p p'} \varepsilon_{p'},
\end{align}
where the \textit{Fock matrix} $F_{p q}$ is defined as
\begin{align} \label{eq:fock}
  F_{p q} \equiv \langle p | \hat{H}_1 | q \rangle + \sum_{r s i'} C_{r i'}^* \langle p r | \hat{H}_2 | q s \rangle C_{s i'}^{},
\end{align}
and $\varepsilon_{p'}$ is a vector of Lagrange multipliers.  The Coulomb and exchange terms are both contained in the second term of Eq.\ \eqref{eq:fock} due to the use of antisymmetrized matrix elements.

Besides trivial cases, the HF equation is generally solved numerically using an iterative algorithm that alternates between the use of Eq.\ \eqref{eq:fock} and Eq.\ \eqref{eq:hartreefock} to successively refine an initial guess for $C_{p p'}$ until a fixed point (self-consistency) is reached.  For our calculations, we use the identity matrix as the initial guess.  When convergence is too slow, methods such as DIIS \cite{PULAY1980393,JCC:JCC540030413}, Broyden's method \cite{broyden1965class}, or even \textit{ad hoc} linear mixing can improve and accelerate convergence greatly.  For our quantum dot cases, linear mixing was more than adequate.

HF does not provide an exact solution to problems where multi-particle correlations are present even if the single-particle basis is not truncated (infinite in size).  The discrepancy between the HF energy and the exact ground state energy is often referred to as the \textit{correlation energy}.  The focus of post-HF methods such as IM-SRG or CC is to add corrections that recover parts of the correlation energy.

To make use of the HF solution as the reference state for post-HF calculations, we transform the matrix elements via Eqs.\ \eqref{eq:hftransform1} and \eqref{eq:hftransform2}.  In effect, this means we are no longer operating within the harmonic oscillator single-particle basis, but rather a HF-optimized single-particle basis.  However, we will omit the prime symbols as the post-HF methods are generally basis-agnostic.

\subsection{The IM-SRG method}
\label{subsec:imsrgmethod}

\subsubsection{Similarity renormalization group in free space}
\label{subsubsec:srgmethods}

The central theme of similarity renormalization group (SRG) methods is the application of a continuous sequence of unitary transformations on the Hamiltonian to evolve it into a band- or block-diagonal form.  This allows the decoupling of a small, designated \textit{model space} from its larger complementary space.  The problem can thus be truncated to the small model space while preserving a large amount of information about the system.  See for examples Refs.\ \cite{kehrein2006flow,Hergert2016165,lnp936} for derivations and calculational details.

The sequence of transformations is parameterized by a continuous variable $s$ known as the \textit{flow parameter}.  Without loss of generality, we can define $s = 0$ to be the beginning of this sequence, thus $\hat{H}(0)$ is simply the original Hamiltonian.  At any value of $s$, the evolving Hamiltonian $\hat{H}(s)$ is related to the original Hamiltonian by
\begin{align*}
  \hat{H}(s) \equiv \hat{U}(s) \hat{H}(0) \hat{U}^\dagger(s),
\end{align*}
where $U(s)$ is a unitary operator that describes the product of all such transformations since $s = 0$.  Taking the derivative with respect to $s$, we obtain:
\begin{align*}
  \frac{\D}{\D s} \hat{H}(s) = \frac{\D \hat{U}(s)}{\D s} \hat{H}(0) \hat{U}^\dagger(s) + \hat{U}(s) \hat{H}(0) \frac{\D \hat{U}^\dagger (s)}{\D s}.
\end{align*}
If we define an operator $\hat{\eta}(s)$ as
\begin{align} \label{eq:etadefinition}
  \hat{\eta}(s) \equiv \frac{\D \hat{U}(s)}{\D s} \hat{U}^\dagger(s),
\end{align}
we find that it is antihermitian as a result of the unitarity of $\hat{U}(s)$:
\begin{align*}
  \hat{\eta}(s) + \hat{\eta}^\dagger(s)
  = \frac{\D}{\D s} \left(\hat{U}(s) \hat{U}^\dagger(s)\right)
  = 0,
\end{align*}
From this property we can derive a differential equation known as the \textit{SRG flow equation}:
\begin{align} \label{eq:imsrgode}
  \frac{\D \hat{H}(s)}{\D s} = [\hat{\eta}(s), \hat{H}(s)].
\end{align}
This equation allows $\hat{H}(s)$ to be evaluated without explicitly constructing the full transformation $\hat U(s)$.  The focus is instead shifted to the operator $\hat \eta(s)$, the \textit{generator} of the transformation.  When $\hat \eta(s)$ is \emph{multiplicatively} integrated (\textit{product integral}), the full unitary transformation $\hat U(s)$ is recovered:
\begin{align} \label{eq:etaintegral}
  \hat U(s')
  = \lim_{\Delta s \to 0} \prod_{i = 1}^n \mathrm{e}^{\hat{\eta}(s_i) \Delta s},
\end{align}
where $s_i \equiv i \Delta s$, $n \equiv \lfloor s' / \Delta s \rfloor$, and $\lfloor x \rfloor$ denotes the \textit{floor} of $x$.  This is the formal solution to the linear differential equation Eq.\ (\ref{eq:etadefinition}).  The product integral in Eq.\ (\ref{eq:etaintegral}) may also be reinterpreted as \textit{$s$-ordering} \cite{reimann2013quantum} in analogy to time-ordering from quantum field theory.

The power of SRG methods lies in the flexibility of the generator $\hat{\eta}$, which is usually chosen in an $s$-dependent manner.  In particular, it is often dependent on the evolving Hamiltonian $\hat{H}(s)$.  The operator $\hat{\eta}$ determines which parts of the Hamiltonian matrix would become suppressed by the evolution, which are usually considered ``off-diagonal'' in an abstract sense.  The ``off-diagonal'' parts could be elements far away from the matrix diagonal, in which case the evolution drives the matrix towards a band-diagonal form.  Or, the ``off-diagonal'' parts could be elements that couple the ground state from the excited state, in which case the evolution drives the matrix towards a block-diagonal form that isolates the ground state.  Or, the ``off-diagonal'' could be literally the elements that do not lie on the diagonal, in which case the evolution would simply diagonalize the Hamiltonian.  Through different choices of $\hat{\eta}$, the SRG evolution can be controlled and adapted to the features of a particular problem.

\subsubsection{Evolving the flow equation in medium}

The SRG flow equation Eq.\ \eqref{eq:imsrgode} can be solved in the second quantization formalism described in Section \ref{subsec:modelHamiltonian}, where creation and annihilation operators are defined with respect to the physical vacuum state.  However, since the basis of a many-body problem grows factorially with the number of particles and the size of the model space, the applicability of the naive (free-space) SRG method is restricted to comparatively small systems.  A more practical approach is to perform the evolution \textit{in medium}, i.e.\ using a many-body Slater determinant as a reference \cite{kehrein2006flow}.  This gives rise to the IM-SRG method \cite{PhysRevC.85.061304,Hergert2016165,lnp936}.

The Hamiltonian representation in Eq.\ \eqref{eq:second_quantized_hamiltonian} is said to be \textit{normal-ordered}\cite{shavitt2009many} with respect to the (true) vacuum, meaning that for any $k$-body operator, its matrix elements vanish unless the bra and ket states have at least $k$ particles.  It is also possible to normal order with respect to a different state, in which case matrix elements of $k$-body operators vanish unless the bra and ket states have at least $k$ \emph{quasiparticles}, as defined with respect to the reference state.  In single-reference IM-SRG, we assume the reference state to be an $N$-particle Slater determinant $\ket{\Phi} \equiv \ket{i_1 i_2 \ldots i_N}$.  We may then rewrite the Hamiltonian $\hat H$ in a representation normal-ordered with respect to the reference state (``Fermi vacuum'') $\ket{\Phi}$:
\begin{align}
  \hat{H} = H_\varnothing + \sum_{p q} H_{p q} \normord{\hat{a}_p^\dagger \hat{a}_q^{}} + \frac{1}{4} \sum_{p q r s} H_{p q r s} \normord{\hat{a}_p^\dagger \hat{a}_q^\dagger \hat{a}_s^{} \hat{a}_r^{}},
  \label{eq:normordhamiltonian}
\end{align}
where
\begin{align*}
  H_\varnothing &\equiv \sum_i \langle i | \hat{H}_1 | i \rangle + \frac{1}{2} \sum_{i j} \langle i j | \hat{H}_2 | i j \rangle, \\
  H_{p q} &\equiv \langle p | \hat{H}_1 | q \rangle + \sum_i \langle p i | \hat{H}_2 | q i \rangle, \\
  H_{p q r s} &\equiv \langle p q | \hat{H}_2 | r s \rangle.
\end{align*}
The colons in $\normord{\hat{a}_p^\dagger \hat{a}_q^{}}$ and $\normord{\hat{a}_p^\dagger \hat{a}_q^\dagger \hat{a}_s^{} \hat{a}_r^{}}$ denote strings of creation and annihilation operators normal-ordered with respect to $\ket{\Phi}$.

If $\Phi$ is a state optimized by the HF method, then $H_\varnothing$ is simply the HF energy $E_\Phi$ in Eq.\ \eqref{eq:hfenergy} and $H_{p q}$ is the Fock matrix $F_{p q}$ in Eq.\ \eqref{eq:fock}.  The operator $\hat H$ in Eq.\ \eqref{eq:normordhamiltonian} is completely equivalent to $\hat H$ in Eq.\ \eqref{eq:second_quantized_hamiltonian}.  The only difference is that the meaning of a ``$k$-body'' operator has been redefined with respect to $\ket{\Phi}$ rather than the vacuum state $\ket{}$, causing matrix elements to be reshuffled among the the $k$-body components.  This makes a critical difference when operator expressions are \emph{truncated}, i.e.\ higher-body operators discarded from the computation for efficiency reasons.  By normal-ordering with respect to an approximate state $\ket{\Phi}$, we reshuffle higher-body contributions into the lower-body terms, significantly reducing the importance of the higher-body operators.

Higher-body operators arise from integrating the flow equations of Eq.\ (\ref{eq:imsrgode}), which is one of the major challenges of the SRG method.  With each evaluation of the commutator, the Hamiltonian gains terms of higher order, and these induced contributions will in subsequent integration steps feed back into terms of lower order.  Thus, the higher-body contributions are not irrelevant to the final solution even if only the ground state energy (zero-body component) is of interest.

Computationally, higher-body terms rapidly become unfeasible to handle: the amount of memory required to store a $k$-body operator grows exponentially with $k$.  Moreover, the flow equations are capable of generating an infinite number of higher-body terms as the Hamiltonian evolves.  Thus, to make the method tractable, the IM-SRG flow equations must be closed by truncating the equations to a finite order.

In this paper, we truncate both $\hat{H}$ and $\hat{\eta}$ at the two-body level, leading to an approach known as IM-SRG(2).  This normal-ordered two-body approximation appears to be sufficient in many cases and has yielded excellent results for several nuclei \cite{PhysRevLett.106.222502,PhysRevLett.109.052501,Hergert2016165}.  The IM-SRG(2) ground state energy is third-order exact, with the most prominent error terms corresponding to fourth order energy diagrams with intermediate triples, and asymmetric intermediate quadruples.  For a wide variety of systems across nuclear physics and quantum chemistry, reincorporation of the triples adds attraction to the system, while that of the asymmetric quadruples adds repulsion.  Because of this cancellation of errors, the IM-SRG(2) tends to track with higher-order methods such as CCSD(T).

Despite the fortuitous result, the cancellation is nonetheless accidental.  A more controlled restoration of the missing error terms can be performed perturbatively.  In Ref.\ \cite{morris2016thesis}, the difference between IM-SRG(2) and perturbative approximations to IM-SRG(3) was investigated in nuclei, the homogeneous electron gas, and several molecules.  In nuclei, differences in binding energies were typically on the order of hundreds of \si{keV}, whereas the binding energies themselves are on the order of \SI{8}{MeV} per nucleon.  This is within the tolerances prescribed by the accuracy of contemporary inter-nucleon interactions.  For electronic systems, the differences were typically less than a few millihartrees.

Note that the loss of three- and higher-body terms (\textit{operator truncation}) is only one out of the two sources of error in this method.  The other source of error is due to the \textit{basis truncation}, a concern for any approach that relies on a finite single-particle basis, including HF, IM-SRG, CC, and many others.  This second source of error can be reduced by increasing the size of the basis at the expense of greater computational effort, albeit the cost increases much less rapidly in this direction.  The CPU cost of IM-SRG methods is polynomial with respect to $N_{\mathrm{b}}$, the number of states in the single-particle basis.  For IM-SRG(2) in particular, the CPU cost scales roughly as $\mathcal{O}(N_{\mathrm{b}}^6)$.

With the operator truncation, the generator $\hat{\eta}$ can be written as a generic 2-body operator:
\begin{align*}
\hat{\eta} = \sum_{p q} \eta_{p q} \normord{\hat a_p^\dagger \hat a_q} +
\frac{1}{4} \sum_{p q r s}\eta_{p q r s} \normord{\hat a_p^\dagger \hat a_q^\dagger \hat a_s \hat a_r},
\end{align*}
where $\eta_{p q}$ and $ \eta_{p q r s}$ respectively are its one- and two-body matrix elements normal ordered with respect to $\Phi$, subject to the antihermittivity constraint.

By expanding the commutator in Eq.\ \eqref{eq:imsrgode} and discarding the three-body term, we obtain the matrix-element form of the IM-SRG(2) flow equation:
\begin{align}
    \frac{\D H}{\D s} = C(\eta, H), \label{eq:flowmxe}
\end{align}
where
\begin{align}
  C(A, B) &\equiv 2 \antisymm_{A B} D(A, B), \label{eq:commutmxe} \\
  \antisymm_{x y} f(x, y) &\equiv \frac{1}{2} \bigl(f(x, y) - f(y, x)\bigr),
\end{align}
\begin{align}
  D_\varnothing(A, B)
  &\equiv
    \sum_{i a} A_{i a} B_{a i}
    + \frac{1}{4} \sum_{i j a b} A_{i j a b} B_{a b i j},
    \label{eq:flow0} \\
  D_{p q}(A, B)
  &\equiv
    \sum_r A_{p r} B_{r q}
    - \frac{1}{2} \sum_{i j a} A_{i j a q} B_{a p i j}
    + \frac{1}{2} \sum_{i a b} A_{i p a b} B_{a b i q}
    \notag \\
  &\hphantom{=}
    + \sum_{i a} \left(
    A_{i a} B_{a p i q}
    + A_{i p a q} B_{a i}
    \right),
    \label{eq:flow1} \\
  D_{p q r s}(A, B)
  &\equiv
    4 \antisymm_{p q} \antisymm_{r s} \sum_{i a} A_{i p a r} B_{a q i s}
    + \frac{1}{2} \sum_{i j} A_{i j r s} B_{p q i j}
    + \frac{1}{2} \sum_{a b} A_{p q a b} B_{a b r s}
    \notag \\
  &\hphantom{=}
    + 2 \sum_t \left(
    \antisymm_{p q} A_{q t} B_{p t r s}
    + \antisymm_{r s} A_{p q r t} B_{t s}
    \right).
    \label{eq:flow2}
\end{align}

\begin{figure}
\includegraphics{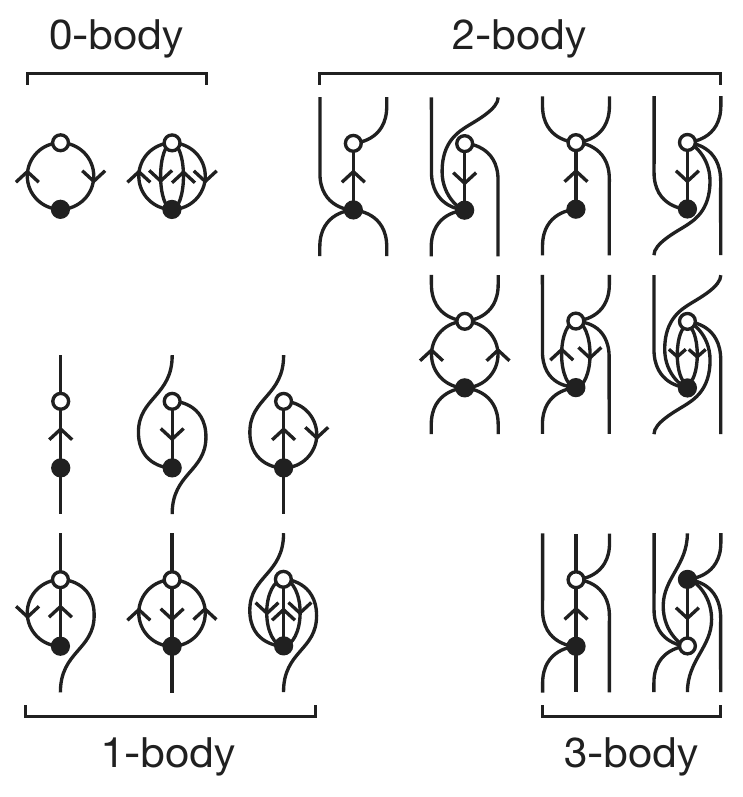}
\caption{(Color online) Diagrammatic representation of $D(\circ, \bullet)$ of Eq.\ \eqref{eq:flow0}, Eq.\ \eqref{eq:flow1}, and Eq.\ \eqref{eq:flow2}, where open circles represent $A$ and filled circles represent $B$.  The two-body vertices are implicitly antisymmetric (i.e.\ \textit{Hugenholtz diagrams} \cite{HUGENHOLTZ1957481}, as described in \cite{shavitt2009many} \S 4.4.3).  Internal directed lines (bound/dummy indices) that point upward denote summations over unoccupied states, whereas lines that point downward denote summations over occupied states.  External undirected lines denote unconstrained free indices.  In IM-SRG(2), the three-body diagrams are not included.}
\label{fig:diagrams-imsrg}
\end{figure}

The equations can be derived directly from the anticommutation relations of the creation and annihilation operators, but it is more convenient to use Wick's theorem \cite{PhysRev.80.268} or, even more efficiently, diagrammatic techniques \cite{shavitt2009many} to arrive at the results.  In particular, Fig.\ \ref{fig:diagrams-imsrg} shows the diagrammatic form of Eq.\ \eqref{eq:flow0}, Eq.\ \eqref{eq:flow1}, and Eq.\ \eqref{eq:flow2}.

The commutator in the flow equations Eq.\ \eqref{eq:imsrgode} ensures that the evolved state $\hat U(s) \ket{\Phi}$ consists of \emph{linked diagrams} only \cite{shavitt2009many}.  This indicates that IM-SRG is a size-extensive \cite{ISI:A1981MN73700014} method by construction, even if the operators are truncated.

An accurate and robust ODE solver is required to solve Eq.\ \eqref{eq:imsrgode}.  In particular, the solver must be capable of handling the stiffness that often arises in such problems.  For our numerical experiments, we used a high-order ODE solver algorithm by L.\ F.\ Shampine and M.\ K.\ Gordon \cite{shampine1975computer}, which is a multistep method based on the implicit Adams predictor-corrector formulas.  Its source code is freely available \cite{odesolver}.

With an appropriate choice of the generator $\hat{\eta}$, the evolved state $\hat U(s) \ket{\Phi}$ will gradually approach a more ``diagonal'' form.  If the ``diagonal'' form decouples the ground state from the excited states, then $\hat U(\infty) \ket{\Phi}$ would yield the exact ground state solution of the problem if no operator or basis truncations are made.  In particular, $H_\varnothing(\infty)$ would be the exact ground state energy.

The original choice of generator suggested by Wegner \cite{Wegner200177} reads
\begin{align*}
  \hat{\eta}^{\text{Wg}}
  = [\hat{H}^{\text{d}}, \hat{H} - \hat{H}^{\text{d}}]
  = [\hat{H}^{\text{d}}, \hat{H}],
\end{align*}
where $\hat{H}^{\text{d}}$ denotes the ``diagonal'' part of the Hamiltonian and $\hat{H} - \hat{H}^{\text{d}}$ denotes the ``off-diagonal'' part.  This is in the abstract sense described at the end of Section \ref{subsubsec:srgmethods}.

Since $\hat{\eta}^{\text{Wg}}$ is a commutator between two Hermitian operators, it is antihermitian as required for a generator.  Additionally, it can be shown that the commutator has the property of suppressing off-diagonal matrix elements as the state evolves via the flow equation \cite{kehrein2006flow}, as we would like.  Matrix elements ``far'' from the diagonal -- i.e.\ where the Hamiltonian couples states with large energy differences -- are suppressed much faster than those ``close'' to the diagonal.

There exist several other generators in literature.  One choice, proposed by White \cite{White:cond-mat0201346}, makes numerical approaches much more efficient.  The problem with the Wegner generator is the widely varying decaying speeds of the Hamiltonian matrix elements.  Terms with large energy separations from the ground state are suppressed initially, followed by those with smaller energy separations.  This leads to stiffness in the flow equation, which in turn causes numerical difficulties when solving the set of coupled differential equations.

The White generator takes an alternative approach, which is well suited for problems where one is mainly interested in the ground state of a system.  Firstly, instead of driving all off-diagonal elements of the Hamiltonian to zero, the generator focuses exclusively on those that are coupled to the reference state $\Phi$ so as to decouple the reference state from the remaining Hamiltonian.  This reduces the amount of change done to the Hamiltonian, reducing the accuracy lost from the operator truncation.  Secondly, the rate of decay in Hamiltonian matrix elements are approximately normalized by dividing the generator matrix elements by an appropriate factor.  This ensures that the affected elements decay at approximately the same rate, reducing the stiffness of the flow equations.

The White generator is explicitly constructed in the following way \cite{PhysRevLett.106.222502,White:cond-mat0201346}.  Let
\begin{align*}
\hat{\eta}^{\text{Wh}} &\equiv \hat{\eta}' - \hat{\eta}'{}^\dagger,
\end{align*}
where
\begin{align*}
\eta'_{a i} &\equiv \frac{H_{a i}}{\tilde{\Delta}_{a i}}, &
\eta'_{a b i j} &\equiv \frac{H_{a b i j}}{\tilde{\Delta}_{a b i j}},
\end{align*}
and the Epstein--Nesbet energy denominators $\tilde{\Delta}$ \cite{shavitt2009many} are defined as
\begin{align}
\tilde{\Delta}_{a i} &\equiv \Delta_{a i} - H_{a i a i}, \notag \\
\tilde{\Delta}_{a b i j} &\equiv \Delta_{a b i j} + w_{a b i j}, \notag \\
\Delta_{p_1 \ldots p_k q_1 \ldots q_k} &\equiv \sum_{i = 1}^k (H_{p_i p_i} -  H_{q_i q_i}), \label{eq:moellerplessetdenominator} \\
w_{a b i j}
  &\equiv H_{a b a b} - H_{a i a i} - H_{b i b i} \notag \\
  &\qquad + H_{i j i j} - H_{a j a j} - H_{b j b j}. \notag
\end{align}
It is also possible \cite{Hergert2016165} to use $\Delta$ (also defined above), the same energy denominators from M\o ller--Plesset perturbation theory, which leads to a slightly different variant of the White generator.  For our calculations, we use exclusively Epstein--Nesbet denominators.

Compared to the Wegner generator, where the derivatives of the final flow equations contain cubes of the Hamiltonian matrix elements (i.e.\ each term contains a product of 3 one-body and/or two-body matrix elements), the elements in White generators contribute only linearly.  This reduces the stiffness in the differential equation, providing a net increase in computational efficiency as stiff ODE solvers tend to be slower and consume more memory.

Lastly, we note that from the above discussion it is clear that the IM-SRG is closely related to several other well-known methods of quantum chemistry such as coupled cluster theory \cite{shavitt2009many}, canonical transformation theory \cite{White:cond-mat0201346,CTreview}, the irreducible (or anti-Hermitian) contracted Schr\"odinger equation approach \cite{Mazziotti1,Mazziotti2}, and the driven similarity renormalization group method \cite{Evangelista}.  We refer the reader to Ref.\ \cite{HeikoReview} for a discussion of similarities with and differences from these approaches.

\subsection{Coupled cluster theory}
\label{subsec:cctheory}

Coupled cluster (CC) theory is based on expressing the $N$-particle correlated wave function $\ket{\Psi}$ using the exponential ansatz,
\begin{align*}
  \ket{\Psi} = \E^{\hat{T}} \ket{\Phi},
\end{align*}
where $\ket{\Phi}$ is the reference state as before.  The cluster operator $\hat{T} \equiv \hat{T}_1 + \hat{T}_2 + \cdots + \hat{T}_N$, is composed of $k$-particle $k$-hole excitation operators, $\hat{T}_k$,
\begin{align} \label{eq:cc_amps}
  \hat{T}_k \equiv \left(\frac{1}{k!}\right)^2 \sum_{\substack{a_1 \ldots a_k \\ i_1 \ldots i_k}} t_{a_1 \ldots a_k i_1 \ldots i_k} \normord{\hat{a}_{a_1}^\dagger \ldots \hat{a}_{a_k}^\dagger \hat a_{i_k}^{} \ldots \hat{a}_{i_1}^{}},
\end{align}
where the unknown matrix elements, $t_{a_1 \ldots a_k i_1 \ldots i_k}$, are known as \textit{cluster amplitudes} \cite{shavitt2009many}.

Using the CC ansatz, the Schr\"odinger equation,
\begin{align} \label{eq:cc_schrodeq}
  \hat{H} \E^{\hat{T}} \ket{\Phi} = E \E^{\hat{T}} \ket{\Phi},
\end{align}
can be rewritten by left-multiplying by $\bra{\Phi} \E^{-T}$ as,
\begin{align*}
  \bra{\Phi} \bar{H}_{\mathrm{CC}} \ket{\Phi} = E,
\end{align*}
where we define a \textit{coupled cluster effective Hamiltonian},
\begin{align} \label{eq:cc_heff0}
  \bar{H}_{\mathrm{CC}} \equiv \E^{-\hat{T}} \hat{H} \E^{\hat{T}},
\end{align}
in which the wave operator, $\E^{\hat{T}}$, acts as a similarity transform on the Hamiltonian in the same way that $\hat{U}(s)$ acts to transform the Hamiltonian in SRG methods.  An important difference, however, is that the wave operator in CC, which contains no de-excitations, is not unitary, and thus $\bar{H}$ is not Hermitian.

The effective Hamiltonian in Eq.\ \eqref{eq:cc_heff0} can be rewritten with commutators according to the Baker--Campbell--Hausdorff expansion as,
\begin{align*}
  \bar{H}_{\mathrm{CC}} = \hat{H} + [\hat{H}, \hat{T}] + \frac{1}{2!} [[\hat{H}, \hat{T} ], \hat{T}] + \frac{1}{3!} [[[\hat{H}, \hat{T}], \hat{T}], \hat{T}] + \frac{1}{4!} [[[[\hat{H}, \hat{T}], \hat{T}], \hat{T}], \hat{T}],
\end{align*}
which terminates at four-nested commutators due to the two-body nature of the interaction.  Like with IM-SRG, this commutator expression ensures that CC is size-extensive and contains only connected terms.  In addition, because $\hat{T}$ is an excitation operator, terms of the form $\hat{T} \hat{H}$ are disconnected and thus vanish \cite{shavitt2009many}.  Therefore the CC effective Hamiltonian can be further reduced to
\begin{align} \label{eq:cc_heff1}
  \bar{H}_{\mathrm{CC}} = \bigl\{\hat{H} \E^{\hat{T}}\bigr\}_{\mathrm{c}},
\end{align}
where the subscript ``$\mathrm{c}$'' indicates that only connected terms are used.

\begin{figure}
  \includegraphics{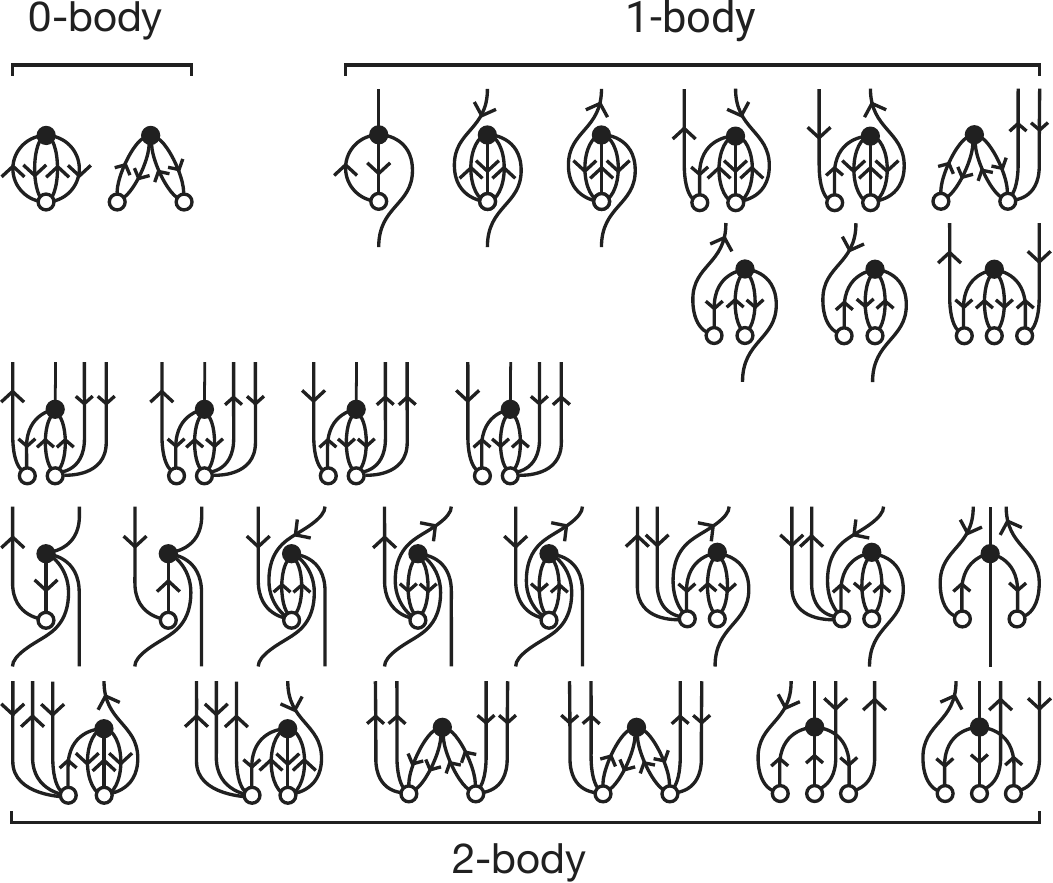}
  \caption{(Color online) Diagrammatic representation of $\bar{H}$ of Eq.\ \eqref{eq:cc_heff1}, excluding terms involving the one-body interaction $\hat{H}_1$ and first-order terms involving only the bare Hamiltonian. Open circles represent the excitation cluster operators $\hat{T}_1$ and $\hat{T}_2$, and filled circles represent the two-body interaction $\hat{H}_2$.  As before, the diagrams are implicitly antisymmetrized (Hugenholtz diagrams).  Lines connected to $\hat{T}$ are always directed upward because they represent an excitation operator while the directions of external lines connected to $\hat{H}_2$ are unconstrained. }
  \label{fig:diagrams-ccsd}
\end{figure}

In practice, the cluster operator $\hat{T}$ must be truncated for calculations to be computationally feasible.  In this work, we use only single and double excitations,
\begin{align*}
  \hat{T} = \hat{T}_1 + \hat{T}_2.
\end{align*}
This is known as coupled cluster with singles and doubles (CCSD), with an asymptotic computational cost that scales like IM-SRG(2).  This truncation has been successfully applied to many problems in quantum chemistry \cite{RevModPhys.79.291} and nuclear physics \cite{2014RPPh...77i6302H}.  In addition, we also truncate the three-body effective Hamiltonian terms that are induced by the similarity transformation.  Fig.\ \ref{fig:diagrams-ccsd} shows the diagrammatic representation of Eq.\ \eqref{eq:cc_heff1} in CCSD.

The unknown cluster amplitudes in CCSD, $t_{a i}$ and $t_{a b i j}$, are calculated by left-multiplying Eq.\ \eqref{eq:cc_schrodeq} by $\bra{\Phi} \normord{\hat a_i^\dagger \hat a_a} \E^{-\hat{T}}$ and $\bra{\Phi} \normord{\hat a_i^\dagger \hat a_j^\dagger \hat a_b \hat a_a} \E^{-\hat{T}}$, respectively,
\begin{align} \label{eq:ccsd1}
  \bra{\Phi} \normord{\hat a_i^\dagger \hat a_a} \bar{H}_{\mathrm{CC}} \ket{\Phi} &= 0, \\
  \bra{\Phi} \normord{\hat a_i^\dagger \hat a_j^\dagger \hat a_b \hat a_a}\bar{H}_{\mathrm{CC}} \ket{\Phi} &= 0. \notag
\end{align}
After the Fock matrix has been diagonalized, the diagonal components of Eq.\ \eqref{eq:ccsd1} can be separated and, after expanding the exponent in Eq.\ \eqref{eq:cc_heff1}, the non-vanishing terms of the CCSD amplitude equations become,
\begin{gather} \label{eq:ccsd2}
  \bra{\Phi} \normord{\hat a_i^\dagger \hat a_a} \hat{H}_2 \bigl\{\hat{T}_1 + \hat{T}_2 + \hat{T}_1\hat{T}_2 + \frac{1}{2!} \hat{T}_1^{2} + \frac{1}{3!} \hat{T}_1^{3}\}_{\mathrm{c}} \ket{\Phi} = \Delta_{i a} t_{a i} \\
  \bra{\Phi} \normord{\hat a_i^\dagger \hat a_j^\dagger \hat a_b \hat a_a} \hat{H}_2 \bigl\{1 + \hat{T}_1 + \hat{T}_2 + \frac{1}{2} \hat{T}_1^{2} + \hat{T}_1\hat{T}_2 + \frac{1}{2!} \hat{T}_2^{2} + \frac{1}{3!} \hat{T}_1^{3} + \frac{1}{2!} \hat{T}_1^{2} \hat{T}_2 + \frac{1}{4!} \hat{T}_1^{4}\bigr\}_{\mathrm{c}} \ket{\Phi} \notag \\
= \Delta_{i j a b} t_{a b i j} \notag
\end{gather}
where $\Delta$ are the M\o ller--Plesset denominators from Eq.\ \eqref{eq:moellerplessetdenominator}.  As usual, these non-linear equations are solved using an iterative procedure where the cluster amplitudes on the right-hand side of Eq.\ \eqref{eq:ccsd2} are updated by calculating the terms on the left-hand side until a fixed point is reached.  Like the HF iterative procedure, employing convergence acceleration techniques can reduce the number of CC iterations required.

\subsection{Quasigenerate perturbation theory}
\label{subsec:selfenergy}

The IM-SRG method provides a means to calculate the ground state energy of any system that is reasonably approximated by a single Slater determinant.  This works well for closed-shell systems, but it does not provide a direct means to obtain the ground state energy of open-shell systems.  While there exist more complicated multi-reference approaches to IM-SRG that seek to tackle the general problem \cite{Hergert2016165}, we opted to use a perturbative approach, which is simple, inexpensive, and as we shall see from the results, quite effective for many problems.

Quasidegenerate perturbation theory (QDPT) is an extension to the traditional perturbation theory framework that incorporates multiple reference states.  It provides us with a simple means to extract ground state energies of open-shell systems that are only a few particles away from a closed-shell system.  In particular, it allows us to calculate \textit{addition energies} $\varepsilon_a$ and \textit{removal energies} $\varepsilon_i$ of such systems, which we define as:
\begin{align}
  \varepsilon_a &\equiv E_{\Phi_a} - E_{\Phi}, \\
  \varepsilon_i &\equiv E_{\Phi} - E_{\Phi_i},
\end{align}
where $i$ is restricted to labels of occupied states, $a$ is restricted to labels of unoccupied states, $|\Phi_i\rangle \equiv \hat{a}_i^{} |\Phi\rangle$, and $|\Phi_a\rangle \equiv \hat{a}_a^\dagger |\Phi\rangle$.

In QDPT, solutions of an approximate one-body Hamiltonian $\hat{H}_1$ form the basis of the model space.  One begins by assuming the existence of an operator $\hat{\Omega}$, known as the \textit{wave operator}, that maps some set of states $\tilde \Psi^{\mathrm{o}}_u$ within the model space to the exact ground state $\Psi_u$:
\begin{align} \label{eq:omega-condition1}
  \Psi_u = \hat \Omega \tilde \Psi^{\mathrm{o}}_u.
\end{align}
The states $\tilde \Psi^{\mathrm{o}}_u$ consist of some mixture of the eigenstates $\Psi^{\mathrm{o}}_{u'}$ of the approximate Hamiltonian $\hat{H}_1$.

There is some freedom in the choice of the wave operator $\hat \Omega$.  We assume it has the following form:
\begin{align} \label{eq:omega-condition2}
  \hat \Omega = \hat P + \hat Q \hat \Omega \hat P,
\end{align}
where $\hat P$ projects any state into the model space and $\hat Q$ is the complement of $\hat P$.  This entails that the exact states $\Psi_u$ are no longer normalized but instead satisfy the so-called intermediate normalization: $\langle \Psi_u | \tilde \Psi^{\mathrm{o}}_u \rangle = 1$.

Making use of the assumptions in Eq.\ \eqref{eq:omega-condition1} and Eq.\ \eqref{eq:omega-condition2}, one can derive from the Schr\"odinger equation the generalized Bloch equation, the principal equation of QDPT:
\begin{gather*}
  [\hat \Omega, \hat{H}_1] =
  (1 - \hat \Omega) \hat V \Omega,
\end{gather*}
where $\hat V \equiv \hat H - \hat{H}_1$ is the perturbation.  The commutator on the left may be ``inverted'' using the resolvent approach (\cite{shavitt2009many}, p.\ 50), resulting in:
\begin{align*}
  \hat Q \Omega \hat P_u =
  \hat R_u (1 - \hat \Omega) \hat V \Omega \hat P_u,
\end{align*}
where $\hat R_u \equiv \hat Q (E_u - \hat Q \hat{H}_1 \hat Q)^{-1} \hat Q$ defines the resolvent and $\hat P_u$ is the projection operator that projects any state onto $\Psi^{\mathrm{o}}_u$.  As is standard in perturbation theory, we now assume $\hat \Omega$ can be expanded as a series of terms of increasing order, as quantified by the power of the perturbation $\hat V$:
\begin{align*}
  \hat \Omega = \hat P +
  \hat Q\bigl(\hat \Omega^{(1)} + \hat \Omega^{(2)} + \cdots\bigr) \hat P.
\end{align*}
This leads to a recursion relation of $\hat \Omega$ that enables $\hat \Omega$ to be calculated up to any order, at least in principle.  Up to third order, we have:
\begin{align*}
  &\hat \Omega^{(1)} \hat P_u = \hat R_u \hat V \hat P_u, \\
  &\hat \Omega^{(2)} \hat P_u =
    \hat R_u \biggl(
    \hat V \hat R_u
    - \sum_v \hat R_v \hat V \hat P_v
    \biggr) \hat V \hat P_u, \\
  &\hat \Omega^{(3)} \hat P_u =
    \hat R_u \biggl(
    \hat V \hat R_u \hat V \hat R_u
    - \hat V \hat R_u \sum_v \hat R_v \hat V \hat P_v \\
  &\qquad\qquad
    - \sum_v \hat R_v \hat V \hat P_v \hat V \hat R_u
    - \sum_v \hat R_v \hat V \hat R_v \hat V \hat P_v \\
  &\qquad\qquad
    + \sum_v \hat R_v \sum_w \hat R_w \hat V \hat P_w \hat V \hat P_v
    \biggr) \hat V \hat P_u.
\end{align*}

\begin{figure}
\includegraphics{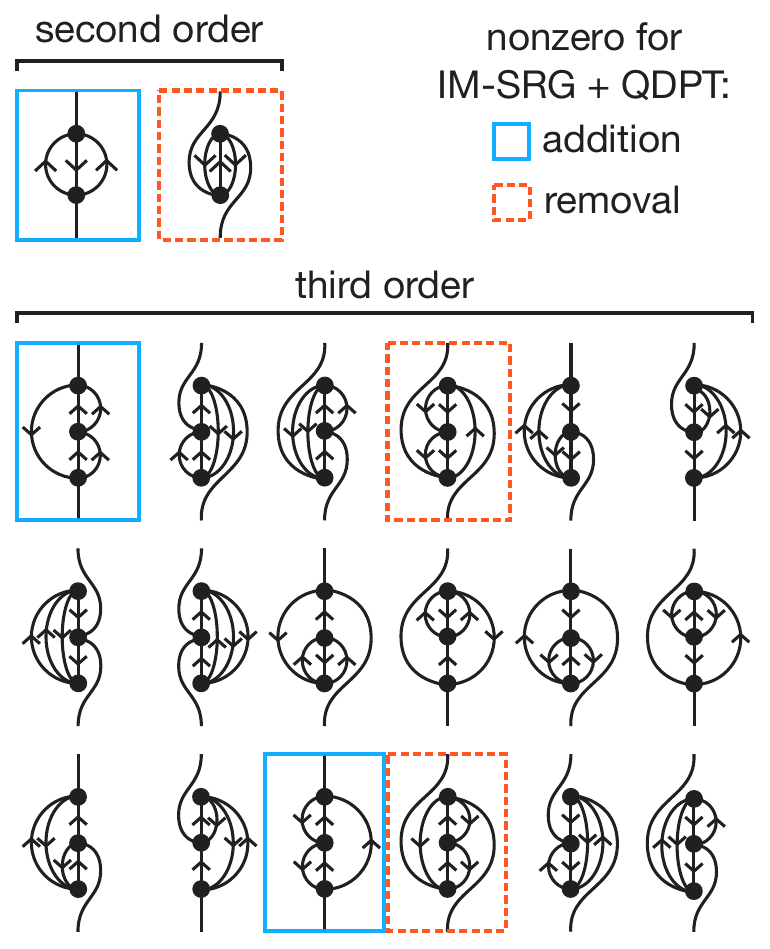}
\caption{(Color online) Diagrammatic form of the second- and third-order QDPT corrections.  The diagrams are implicitly antisymmetrized (Hugenholtz diagrams), but also have implicit denominators as with many-body perturbation theory.  When QDPT is performed on IM-SRG-evolved Hamiltonians, many of the diagrams vanish.  The remaining nonvanishing diagrams for addition energy are highlighted in blue and for removal energy are highlighted in red.}
\label{fig:diagrams-sfe}
\end{figure}

To make the equations more concrete, we further assume that each reference state $\ket{\Psi^{\mathrm{o}}_u} \equiv \ket{\Phi_u}$, i.e.\ each reference state is simply a Slater determinant constructed by adding or removing a single particle $u$ to a closed-shell reference state $\Phi$, which itself may have been obtained earlier from HF and/or IM-SRG.  Thus, the number of reference states for QDPT is equal to the number of particles in either the lowest unfilled shell or the highest filled shell of $\ket{\Phi}$, depending on whether we are considering addition or removal energies.

We can then express the perturbation expansion in terms of summations over matrix elements as we did for the IM-SRG flow equation, again with the aid of Wick's theorem or diagrammatic techniques.  This leads to the following expression for the second-order correction:
\begin{align*}
  \varepsilon_p^{(2)}
  &=
    \sum_{i a b} \frac{|H_{p i a b}|^2}{2 \Delta_{p i a b}}
    - \sum_{i j a} \frac{|H_{i j p a}|^2}{2 \Delta_{i j p a}},
\end{align*}
where the M\o ller--Plesset denominators $\Delta$ are defined in Eq.\ \eqref{eq:moellerplessetdenominator}.  The above expression is depicted in Fig.\ \ref{fig:diagrams-sfe}.  Since there are numerous terms in the third-order correction $\varepsilon_p^{(3)}$, they are listed in diagrammatic form in Fig.\ \ref{fig:diagrams-sfe}.  We will refer to QDPT to third order as ``QDPT3''.

Taking into account the perturbation corrections, one can extract reasonably accurate addition and removal energies for single-particle states near the Fermi level via:
\begin{align*}
  \varepsilon_p = H_{p p} + \varepsilon_p^{(2)} + \varepsilon_p^{(3)}.
\end{align*}

There is some degree of synergy between IM-SRG and QDPT: a generator that decouples the ground state energy will necessarily drive certain classes of matrix elements to zero.  This means certain kinds of vertices in the diagrams become forbidden, reducing the number of nonzero diagrams at third order from 18 to only four.

\subsection{Equations-of-motion methods}

Particle attached and particle removed equations-of-motion (EOM) methods can be coupled with either IM-SRG or CC calculations. The principal idea is that one may construct a ladder operator $\hat{X}$ that promotes the $N$-particle ground state to any state in the $N + 1$ or $N - 1$ spectrum,
\begin{equation}\label{eq:ladder_def}
  \ket{\Psi^{(N \pm 1)}_u}  = \hat{X}^{(N \pm 1)}_u \ket{\Psi^{(N)}_0},
\end{equation}
where $\hat{X}$ is in principle a linear combination of excitation $(+)$ and de-excitation $(-)$ operators that change particle number by one,
\begin{align}
  \label{eq:gen_attached}
  \hat{X}^{(N + 1)}_u &= \sum_a x^{(u, +)}_a  \normord{\hat{a}^\dagger_a} + \frac{1}{2} \sum_{a b i} x^{(u, +)}_{a b i} \normord{\hat{a}^\dagger_a \hat{a}^\dagger_b \hat{a}_i^{}} + \cdots,  \\
  \label{eq:gen_removed}
  \hat{X}^{(N - 1)}_u &= \sum_i x_i^{(u, -)} \normord{\hat{a}_i^{}} + \frac{1}{2} \sum_{i j a} x^{(u, -)}_{a i j} \normord{\hat{a}^\dagger_a \hat{a}_j^{} \hat{a}_i^{}}  + \cdots.
\end{align}
Here, $x^{(u, \pm)}_p$ and $x^{(u, \pm)}_{p q r}$ are the normal-ordered matrix elements of $\hat{X}_u^{(N \pm 1)}$, defined analogously to Eq.\ \eqref{eq:normordhamiltonian}.

Substitution of Eq.\ \eqref{eq:ladder_def} into the energy eigenvalue problem
\begin{gather*}
  \hat H \ket{\Psi_u^{(N \pm 1)}} = E^{(N\pm1)}_u \ket{\Psi_u^{(N \pm 1)}},
\end{gather*}
gives
\begin{equation}\label{eq:EOM}
  [\hat{H}, \hat{X}^{(N \pm 1)}_u] \ket{\Psi^{(N)}_0} = \pm \varepsilon^{(\pm)}_u \hat{X}^{(N \pm 1)}_u \ket{\Psi^{(N)}_0},
\end{equation}
which constitutes a generalized eigenvalue problem for the amplitudes $x$, where $\varepsilon^{(\pm)}_u$ are the single-particle addition ($+$) and removal ($-$) energies. The quality of this calculation depends on the ansatz for the $N$-particle ground state, as well as the systematically improvable truncation on the ladder operators. In this work we include 1p and 2p1h excitations in the $N + 1$ ladder operator and likewise 1h and 2h1p operators for the $N - 1$ ladder operators.

\subsubsection*{Equations-of-motion IM-SRG}

After a single-reference ground state IM-SRG calculation, the Hamiltonian has been rotated such that the reference state is an eigenfunction with corresponding eigenvalue $E^{(N)}_0$, which is the correlated $N$-particle ground state energy. The EOM equation is therefore
\begin{align} \label{eq:EOMIMSRG}
  [\bar{H},\bar{X}^{(N \pm 1)}_u] \ket{\Phi^{(N)}_0} = \pm \varepsilon^{(\pm)}_u \bar{X}^{(N \pm 1)}_u \ket{\Phi^{(N)}_0},
\end{align}
where bars denote rotated operators. Now the reference state is used in place of the bare correlated ground state. The ground state IM-SRG procedure has implicitly re-summed contributions from higher order excitations (3-particle-2-hole, 2-particle-3-hole, 2-particle-3-hole, 4-particle-3-hole, \ldots) into the lower order amplitudes of the ladder operators (1-particle-0-hole, 0-particle-1-hole, 2-particle-1-hole, 1-particle-2-hole).

Despite these gains, the EOM calculation is still a partial diagonalization method, limited by the truncation to 2-particle-1-hole and 1-particle-2-hole operators. We expect $N + 1$ (or $N - 1$) states to be described appropriately by EOM-IM-SRG if their wavefunctions are dominated by 1-particle-0-hole (or 0-particle-1-hole) contributions in the rotated frame. We use partial norms of the EOM ladder operators to estimate these contributions:
\begin{align}
  \label{eq:partial_norms_p}
  n_{\text{1-particle}} &= \sqrt{\sum_a | \bar{x}^{(+)}_a |^2},\\
  \label{eq:partial_norms_h}
  n_{\text{1-hole}} &= \sqrt{\sum_i | \bar{x}^{(-)}_i |^2}.
\end{align}
Large single particle partial norms indicate that the EOM truncation is reasonable for the relevant state. States with lower single particle norms should be treated with a higher EOM approximation, which can be accomplished directly or perturbatively \cite{PhysRevC.95.044304}.

\subsubsection*{Equations-of-motion coupled cluster theory}

Like EOM-IM-SRG, the equations-of-motion technique can be applied after a CC ground-state calculation, by using the CC effective Hamiltonian.  Here, the non-Hermitian nature of $\bar{H}_{\mathrm{CC}}$ becomes apparent.  In this case, in addition to constructing excitation ladder operators Eq.\ \eqref{eq:gen_attached} and Eq.\ \eqref{eq:gen_removed} that correspond to the right-eigenvectors of the generalized eigenvalue problem Eq.\ \eqref{eq:EOMIMSRG}, there exist analogous de-excitation ladder operators, $\hat{L}^{(N + 1)}$ and $\hat{L}^{(N - 1)}$, that correspond to the left-eigenvectors,
\begin{align*}
    \hat{L}^{(N + 1)}_u &= \sum_a l^{(u, +)}_a \normord{\hat{a}_a^{}} + \frac{1}{2} \sum_{i a b} l^{(u, +)}_{i a b} \normord{\hat{a}^\dagger_i \hat{a}_b^{} \hat{a}_a^{}} + \cdots, \\
    \hat{L}^{(N - 1)}_u &= \sum_i l_i^{(u, -)} \normord{\hat{a}^\dagger_i} + \frac{1}{2} \sum_{i j a} l^{(u, -)}_{i j a} \normord{\hat{a}^\dagger_i \hat{a}^\dagger_j \hat{a}_a^{}} + \cdots,
\end{align*}
where $l^{(u, \pm)}_p$ and $l^{(u, \pm)}_{p q r}$ are likewise the normal-ordered matrix elements of $\hat{L}^{(N \pm 1)}_u$.  These left-eigenvectors satisfy the left-eigenvalue problem, with left-eigenvalues $\varepsilon^{(\pm)}_u$, analogous to Eq.\ \eqref{eq:EOMIMSRG},
\begin{align*}
  \bra{\Phi^{(N)}_0} [\bar{H}, \bar{L}^{(N \pm 1)}_u] = \pm \varepsilon^{(\pm)}_u \bra{\Phi^{(N)}_0} \bar{L}^{(N \pm 1)}_u
\end{align*}
and form a bi-orthogonal set with the right-eigenvectors, $\braket{\bar{L}^{(N \pm 1)}_u}{\bar{X}^{(N \pm 1)}_v} = \delta_{u v}$.

In this paper, because the effective Hamiltonian is real, the corresponding left- and right-eigenvalues are equal. In addition, while the the left- and right-eigenvectors are generally not equivalent, the differences in their single-particle Eq.\ \eqref{eq:partial_norms_p} or single-hole Eq.\ \eqref{eq:partial_norms_h} natures are, in practice, not significant.  Therefore, only the right-eigenvectors are used in this paper.

\section{Results}
\label{sec:results}

\subsection{Methodology}
\label{subsec:methodology}

\begin{figure}
  \centering
  \includegraphics{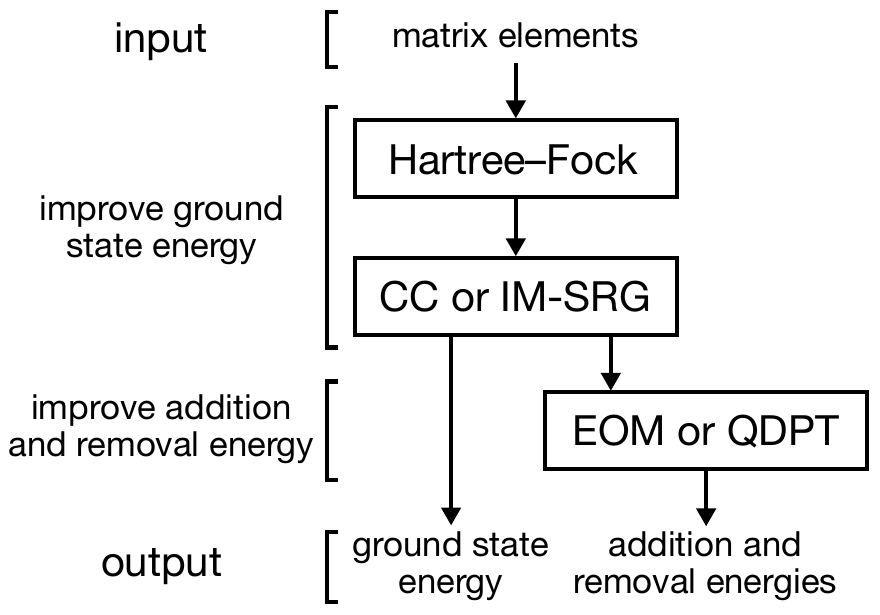}
  \caption{(Color online) A schematic view of the various ways in which many-body methods in this paper could be combined to calculate ground state, addition, and removal energies.}
  \label{fig:methods}
\end{figure}

There is significant flexibility in the application of many-body methods.  The approaches we use are shown in Fig.\ \ref{fig:methods}.  Applying the methods in this order maximizes the benefits of each method: HF acts as an initial, crude procedure to ``soften'' the Hamiltonian, followed by IM-SRG or CC to refine the ground state energy, and then finally QDPT or EOM to refine the addition and removal energies.  We expect single-reference IM-SRG and CC to recover a substantial part of the dynamical correlations, while QDPT and EOM help account for static correlations.

The general process begins with the input matrix elements Eq.\ \eqref{eq:interactionintegral}, computed exactly using the \texttt{OpenFCI} software library \cite{2008arXiv0810.2644K}.  Internally, \texttt{OpenFCI} calculates the interaction integrals in the center-of-mass frame using Gauss--Hermite quadrature and transforms them back into the laboratory frame to produce matrix elements suitable for use in our many-body methods.  In principle, one could also compute Eq.\ \eqref{eq:interactionintegral} analytically \cite{0953-8984-10-3-013}, but the computational cost of the analytic expression grows much more rapidly than \texttt{OpenFCI}'s quadrature-based approach.

Afterward, there are several paths through which one can traverse Fig.\ \ref{fig:methods} to obtain output observables.  We shall primarily focus on the three combinations: (a) HF + IM-SRG(2) + QDPT3, (b) HF + IM-SRG(2) + EOM2, and (c) HF + CCSD + EOM2.

It is possible to omit some steps of the process.  For example, one can omit HF, but continue with the remaining two steps.  While this is doable, from our experience HF significantly improves the results of the later post-HF methods at very low cost compared to the post-HF methods.  Therefore, in practice there is little reason to omit HF.  We will however investigate the effects of removing one or more of the post-HF methods.

Since every calculation in this paper begins with the HF stage, we will not explicitly state ``HF'' unless there is no post-HF method used at all, in which case we write ``HF only''.

All calculations of ground state energy $E_N$ in this paper are restricted to cases where the number of particles $N$ is a magic number, i.e.\ a \textit{closed shell} system (see Fig.\ \ref{fig:shell-structure} for an illustration of the shell structure).  This is a limitation of the many-body methods used in this paper and while there are ways to overcome this limit they are beyond the scope of this paper (see Section \ref{sec:conclusions} for some ideas).  Addition/removal energies $\varepsilon^{(\pm)}$ are similarly restricted in that we only calculate the energy difference between $E_N$ of a closed shell system and $E_{N \pm 1}$ of the same system but with one particle added/removed:
\begin{align*}
  \varepsilon^{(+)} &\equiv E_{(N + 1)} - E_N, \\
  \varepsilon^{(-)} &\equiv E_N - E_{(N - 1)}.
\end{align*}

Ideally, ground state energies should be characterized entirely by the two system parameters $(N, \omega)$, where $N$ is number of particles and $\omega$ is the oscillator frequency.  However, the methods that we study are limited to a \emph{finite} (truncated) basis and the results depend on the level of truncation.  This is characterized by $K$, the total number of shells in the single-particle basis.  Thus, results are generally presented as a graph plotted against $K$.  In Section \ref{subsec:extrapolation} we discuss how to estimate results as $K \to \infty$ (infinite-basis limit) through \textit{extrapolations}.

The addition and removal energies are similar, but they require an additional parameter: the total orbital angular momentum $M_\ell$, defined as the sum of the $m_\ell$ of each particle.  This is due to the presence of multiple states with near-degenerate energies.  For this paper, we will consider exclusively the addition/removal energies with the lowest $|M_\ell|$ subject to the constraint that the particle added/removed lies within the next/last shell.  This means the $N + 1$ states of interest are those with $|M_\ell| = K_{\mathrm{F}} \bmod 2$ (where $\bmod$ stands for the modulo operation) where $K_{\mathrm{F}}$ is the number of occupied shells, while the $N - 1$ states of interest are those with $|M_\ell| = 1 - (K_{\mathrm{F}} \bmod 2)$.

Not all cases are solvable with our selection many-body methods.  Low frequency systems are particularly strenuous for these many-body methods due to their strong correlations, leading to equations that are difficult and expensive to solve numerically.  In the tables, ``n.c.'' marks the cases where IM-SRG(2) or CCSD either diverged or converged extremely slowly.  This also affects the extrapolation results in Section \ref{subsec:extrapolation}, as for consistency reasons we chose to extrapolate only when all five points were available.

Numerical calculations in this paper are performed with a relative precision of about $10^{-5}$ or lower.  This does not necessarily mean the results are as precise as $10^{-5}$, since numerical errors tend to accumulate over the multiple steps of the calculation, thus the precision of the final results is expected to be roughly $10^{-4}$.

\subsection{Comparison between methods}

\subsubsection{Ground state energy}

\begin{table}
  \centering
  \caption{Ground state energy of quantum dots with $N$ particles and an oscillator frequency of $\omega$.  For every row, the calculations are performed in a harmonic oscillator basis with $K$ shells.  The abbreviation ``n.c.'' stands for ``no convergence'': these are cases where IM-SRG(2) or CCSD either diverged or converged extremely slowly.}
  \label{tab:ground}
  \begin{tabular}{S[table-format=2.0]SS[table-format=2.0]S[table-format=3.5]S[table-format=3.5]S[table-format=3.5]S[table-format=3.5]}%
\hline\hline
{$N$} & {$\omega$} & {$K$} & {HF} & {MP2} & {IM-SRG(2)} & {CCSD} \\
\hline
6 & 0.1 & 14 & 3.8524 & 3.5449 & 3.4950 & 3.5831 \\
6 & 0.28 & 14 & 8.0196 & 7.6082 & 7.5731 & 7.6341 \\
6 & 1.0 & 14 & 20.7192 & 20.1939 & 20.1681 & 20.2000 \\
\hline
12 & 0.1 & 16 & 12.9247 & 12.2460 & 12.2215 & 12.3583 \\
12 & 0.28 & 16 & 26.5500 & 25.6433 & 25.6259 & 25.7345 \\
12 & 1.0 & 16 & 66.9113 & 65.7627 & 65.7475 & 65.8097 \\
\hline
20 & 0.1 & 16 & 31.1460 & 29.9674 & 29.9526 & 30.1610 \\
20 & 0.28 & 16 & 63.5388 & 61.9640 & 61.9585 & 62.1312 \\
20 & 1.0 & 16 & 158.0043 & 156.0239 & 156.0233 & 156.1243 \\
\hline
30 & 0.1 & 16 & 62.6104 & 60.8265 & 60.6517 & 61.0261 \\
30 & 0.28 & 16 & 126.5257 & 124.1279 & 124.1041 & 124.3630 \\
30 & 1.0 & 16 & 311.8603 & 308.8611 & 308.8830 & 309.0300 \\
\hline
42 & 0.1 & 20 & 110.7797 & 108.1350 & 108.0604 & 108.5150 \\
42 & 0.28 & 20 & 223.5045 & 219.9270 & 220.0227 & 220.3683 \\
42 & 1.0 & 20 & 547.6832 & 543.2139 & 543.3399 & 543.5423 \\
\hline
56 & 0.1 & 20 & 182.6203 & 179.2370 & {n.c.} & 179.6938 \\
56 & 0.28 & 20 & 363.8784 & 359.1916 & 359.1997 & 359.6744 \\
56 & 1.0 & 20 & 885.8539 & 879.9325 & 880.1163 & 880.3781 \\
\hline\hline
\end{tabular}
\end{table}

\begin{figure}
  \centering
  \includegraphics{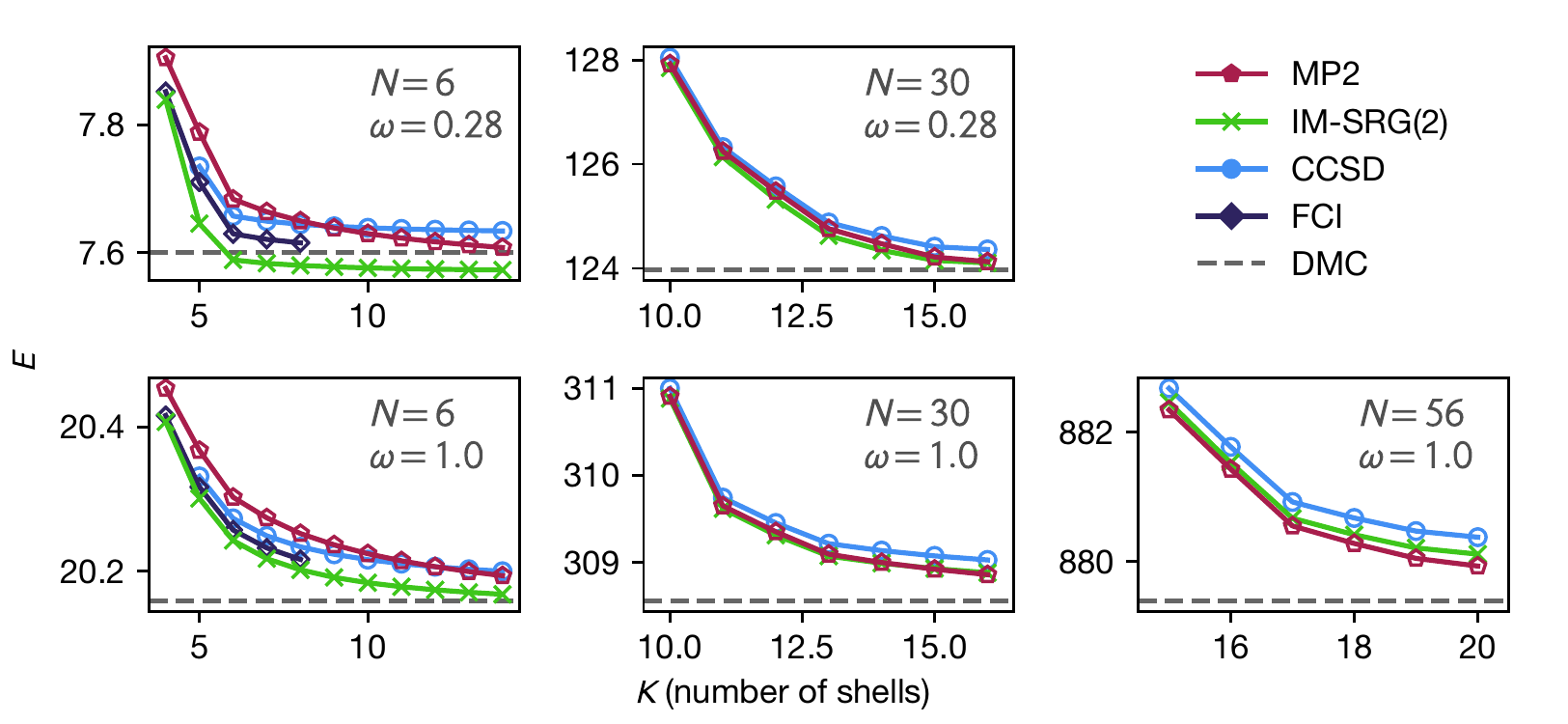}
  \caption{(Color online) Plots of ground state energy of quantum dots with $N$ particles and an oscillatory frequency of $\omega$ against the number of shells $K$.  Since DMC does not utilize a finite basis, the horizontal axis is irrelevant and DMC results are plotted as horizontal lines.}
  \label{fig:gs}
\end{figure}

\begin{table}
  \centering
  \caption{Similar to Table \ref{tab:ground}, this table compares the ground state energies of quantum dots calculated using IM-SRG(2), CCSD, and FCI \cite{olsen2013thesis}.}
  \label{tab:ground-fci}
  \begin{tabular}{S[table-format=2.0]SS[table-format=2.0]S[table-format=3.5]S[table-format=3.5]S[table-format=3.5]}%
\hline\hline
{$N$} & {$\omega$} & {$K$} & {IM-SRG(2)} & {CCSD} & {FCI} \\
\hline
2 & 0.1 & 5 & {n.c.} & 0.4416 & 0.4416 \\
2 & 0.28 & 5 & 0.9990 & 1.0266 & 1.0266 \\
2 & 1.0 & 5 & 3.0068 & 3.0176 & 3.0176 \\
2 & 0.1 & 10 & {n.c.} & 0.4411 & 0.4411 \\
2 & 0.28 & 10 & 0.9973 & 1.0236 & 1.0236 \\
2 & 1.0 & 10 & 2.9961 & 3.0069 & 3.0069 \\
\hline
6 & 0.1 & 8 & 3.4906 & 3.5853 & 3.5552 \\
6 & 0.28 & 8 & 7.5802 & 7.6446 & 7.6155 \\
6 & 1.0 & 8 & 20.2020 & 20.2338 & 20.2164 \\
\hline\hline
\end{tabular}
\end{table}

Fig.\ \ref{fig:gs} and Table \ref{tab:ground} display a selection of ground state energies calculated using HF + IM-SRG(2) and HF + CCSD as described in Section \ref{subsec:methodology}.  We include results from M\o ller--Plesset perturbation theory to second order (MP2), DMC \cite{hoegberget2013thesis}, and FCI \cite{olsen2013thesis} (see Table \ref{tab:ground-fci}) for comparison where available.

We do not include results from ``HF only'' to avoid overshadowing the comparatively smaller differences between the non-HF results in the plots.  Some HF results can be found in Fig.\ \ref{fig:by-freq-10-6-normal} instead.  Generally, the HF ground state energies differ from the non-HF ones by a few to several percent, whereas non-HF energies tend to differ from each other by less than a percent.

With respect to the number of shells, both IM-SRG(2) and CCSD appear to converge slightly faster than second order perturbation theory (MP2), mainly due to the presence of higher order corrections in IM-SRG(2) and CCSD.

There are a few cases where the IM-SRG over-corrects the result, leading to an energy lower than the quasi-exact DMC results.  This is not unexpected given that, unlike the HF results, the IM-SRG method is non-variational in the presence of operator truncations, which in turn results in small unitarity violations.  This over-correction tends to occur when the frequency is low (high correlation), or when \emph{few} particles are involved.

\subsubsection{Addition and removal energies}

\begin{table}
  \centering
  \caption{Addition energy of quantum dot systems.  See Table \ref{tab:ground} for details.}
  \label{tab:add}
  \begin{tabular}{S[table-format=2.0]SS[table-format=2.0]S[table-format=3.5]S[table-format=3.5]S[table-format=3.5]S[table-format=3.5]}%
\hline\hline
{$N$} & {$\omega$} & {$K$} & {HF} & {IM-SRG(2)} & {IMSRG(2)} & {CCSD} \\
{} & {} & {} & {+QDPT3} & {+QDPT3} & {+EOM} & {+EOM} \\
\hline
6 & 0.1 & 14 & 1.2272 & 1.2014 & 1.1809 & 1.1860 \\
6 & 0.28 & 14 & 2.5208 & 2.5003 & 2.4916 & 2.4833 \\
6 & 1.0 & 14 & 6.4804 & 6.4546 & 6.4532 & 6.4453 \\
\hline
12 & 0.1 & 16 & 1.9716 & 1.9248 & 1.9094 & 1.9014 \\
12 & 0.28 & 16 & 3.9901 & 3.9394 & 3.9354 & 3.9205 \\
12 & 1.0 & 16 & 9.9742 & 9.9256 & 9.9274 & 9.9136 \\
\hline
20 & 0.1 & 16 & 2.8000 & 2.7143 & 2.7149 & 2.7040 \\
20 & 0.28 & 16 & 5.6193 & 5.5400 & 5.5409 & 5.5226 \\
20 & 1.0 & 16 & 13.8468 & 13.7799 & 13.7844 & 13.7667 \\
\hline
30 & 0.1 & 16 & 3.7880 & 3.6467 & 3.6536 & 3.6454 \\
30 & 0.28 & 16 & 7.3971 & 7.2719 & 7.2810 & 7.2615 \\
30 & 1.0 & 16 & 17.9948 & 17.9022 & 17.9088 & 17.8875 \\
\hline
42 & 0.1 & 20 & 4.7647 & 4.5751 & 4.5867 & 4.5750 \\
42 & 0.28 & 20 & 9.2798 & 9.1072 & 9.1188 & 9.0963 \\
42 & 1.0 & 20 & 22.4142 & 22.2941 & 22.3012 & 22.2766 \\
\hline
56 & 0.1 & 20 & 6.0029 & {n.c.} & {n.c.} & 5.7661 \\
56 & 0.28 & 20 & 11.3932 & 11.1683 & 11.1813 & 11.1518 \\
56 & 1.0 & 20 & 27.0513 & 26.9033 & 26.9118 & 26.8842 \\
\hline\hline
\end{tabular}
\end{table}

\begin{table}
  \centering
  \caption{Removal energy of quantum dot systems.  See Table \ref{tab:add} for details.}
  \label{tab:rm}
  \begin{tabular}{S[table-format=2.0]SS[table-format=2.0]S[table-format=3.5]S[table-format=3.5]S[table-format=3.5]S[table-format=3.5]}%
\hline\hline
{$N$} & {$\omega$} & {$K$} & {HF} & {IM-SRG(2)} & {IMSRG(2)} & {CCSD} \\
{} & {} & {} & {+QDPT3} & {+QDPT3} & {+EOM} & {+EOM} \\
\hline
6 & 0.1 & 14 & 1.0480 & 0.9500 & 0.9555 & 1.0054 \\
6 & 0.28 & 14 & 2.1171 & 2.0346 & 2.0398 & 2.0782 \\
6 & 1.0 & 14 & 5.2513 & 5.1950 & 5.1970 & 5.2220 \\
\hline
12 & 0.1 & 16 & 1.8003 & 1.6961 & 1.7017 & 1.7503 \\
12 & 0.28 & 16 & 3.6258 & 3.5334 & 3.5366 & 3.5779 \\
12 & 1.0 & 16 & 8.8790 & 8.8104 & 8.8102 & 8.8409 \\
\hline
20 & 0.1 & 16 & 2.6223 & 2.5133 & 2.5184 & 2.5670 \\
20 & 0.28 & 16 & 5.2652 & 5.1639 & 5.1660 & 5.2105 \\
20 & 1.0 & 16 & 12.7991 & 12.7201 & 12.7185 & 12.7546 \\
\hline
30 & 0.1 & 16 & 3.5755 & 3.4445 & 3.4485 & 3.5113 \\
30 & 0.28 & 16 & 7.0306 & 6.9282 & 6.9289 & 6.9785 \\
30 & 1.0 & 16 & 17.0045 & 16.9243 & 16.9215 & 16.9613 \\
\hline
42 & 0.1 & 20 & 4.5209 & 4.3868 & 4.3902 & 4.4451 \\
42 & 0.28 & 20 & 8.8831 & 8.7765 & 8.7766 & 8.8263 \\
42 & 1.0 & 20 & 21.4334 & 21.3453 & 21.3421 & 21.3848 \\
\hline
56 & 0.1 & 20 & 5.6904 & {n.c.} & {n.c.} & 5.6341 \\
56 & 0.28 & 20 & 10.9602 & 10.8471 & 10.8454 & 10.8957 \\
56 & 1.0 & 20 & 26.0952 & 26.0094 & 26.0056 & 26.0507 \\
\hline\hline
\end{tabular}
\end{table}

\begin{figure}
  \centering
  \includegraphics{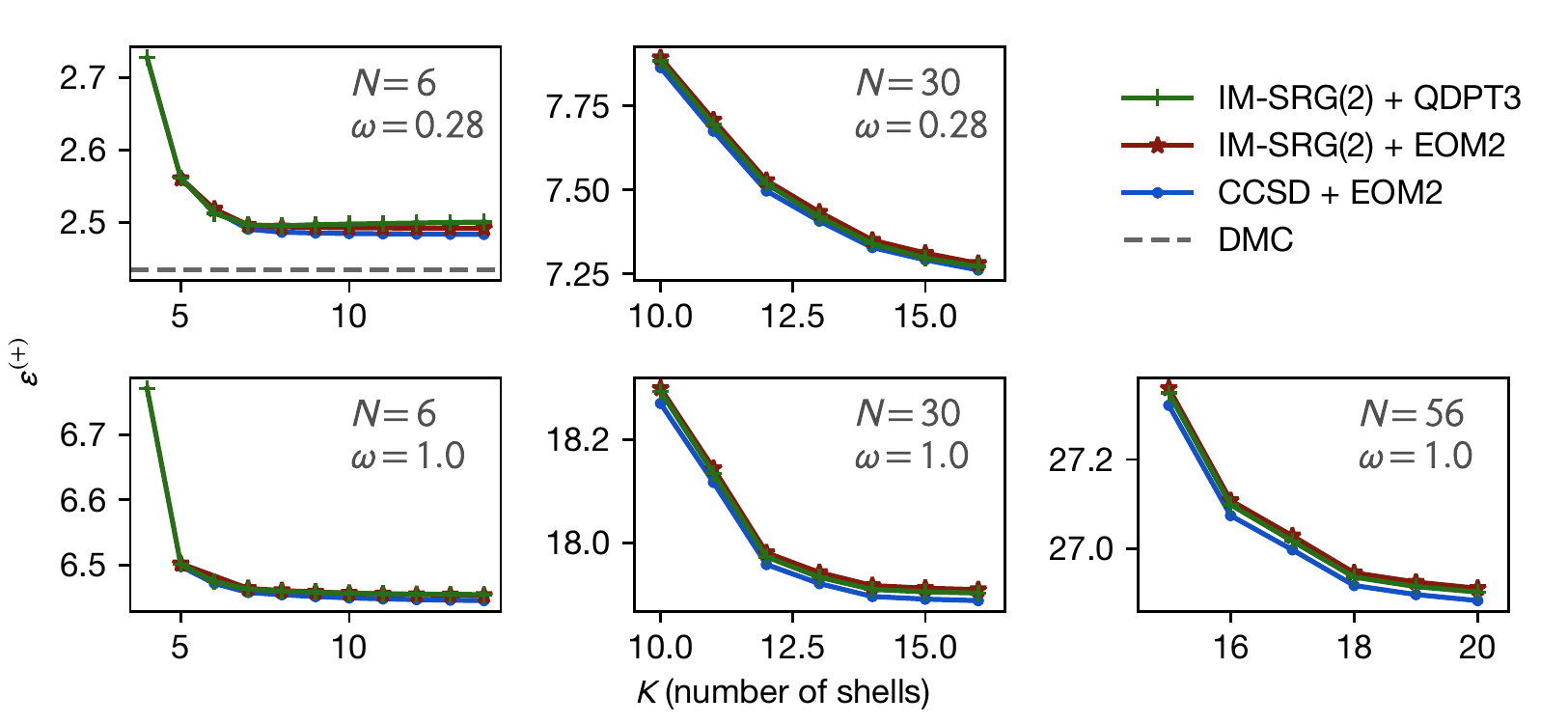}
  \caption{(Color online) Addition energies for a selection of quantum dot parameters.  See Fig.\ \ref{fig:gs} for details.}
  \label{fig:add}
\end{figure}

\begin{figure}
  \centering
  \includegraphics{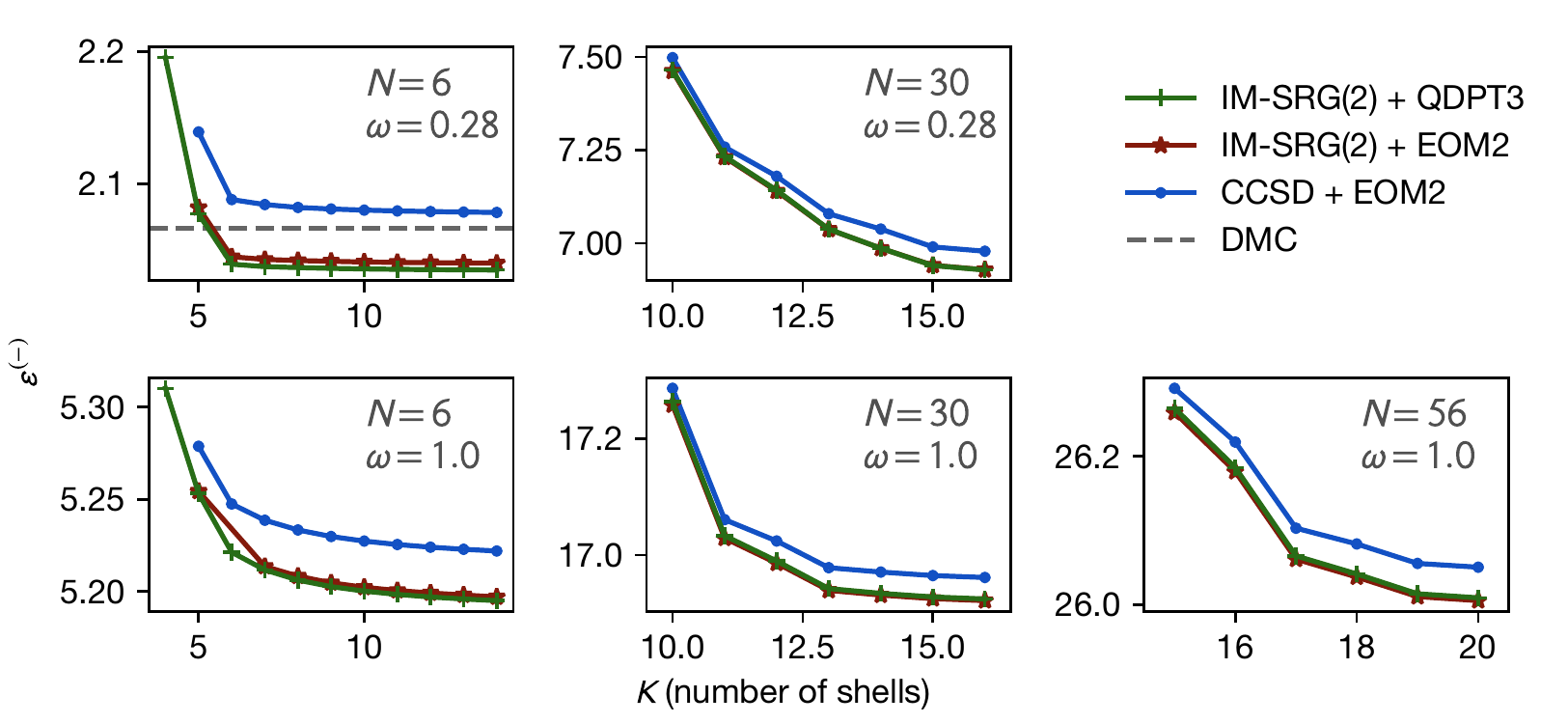}
  \caption{(Color online) Removal energies for a selection of quantum dot parameters.  See Fig.\ \ref{fig:add} for details.}
  \label{fig:rm}
\end{figure}

The results of our addition and removal energy calculations are summarized in Fig.\ \ref{fig:add} and Fig.\ \ref{fig:rm} respectively.  The figures show the the addition/removal energies for using the approaches mentioned in Section \ref{subsec:methodology}.  Where available, results from diffusion Monte Carlo (DMC) \cite{PhysRevB.84.115302} are shown as a dashed line.

\begin{figure}
  \centering
  \includegraphics{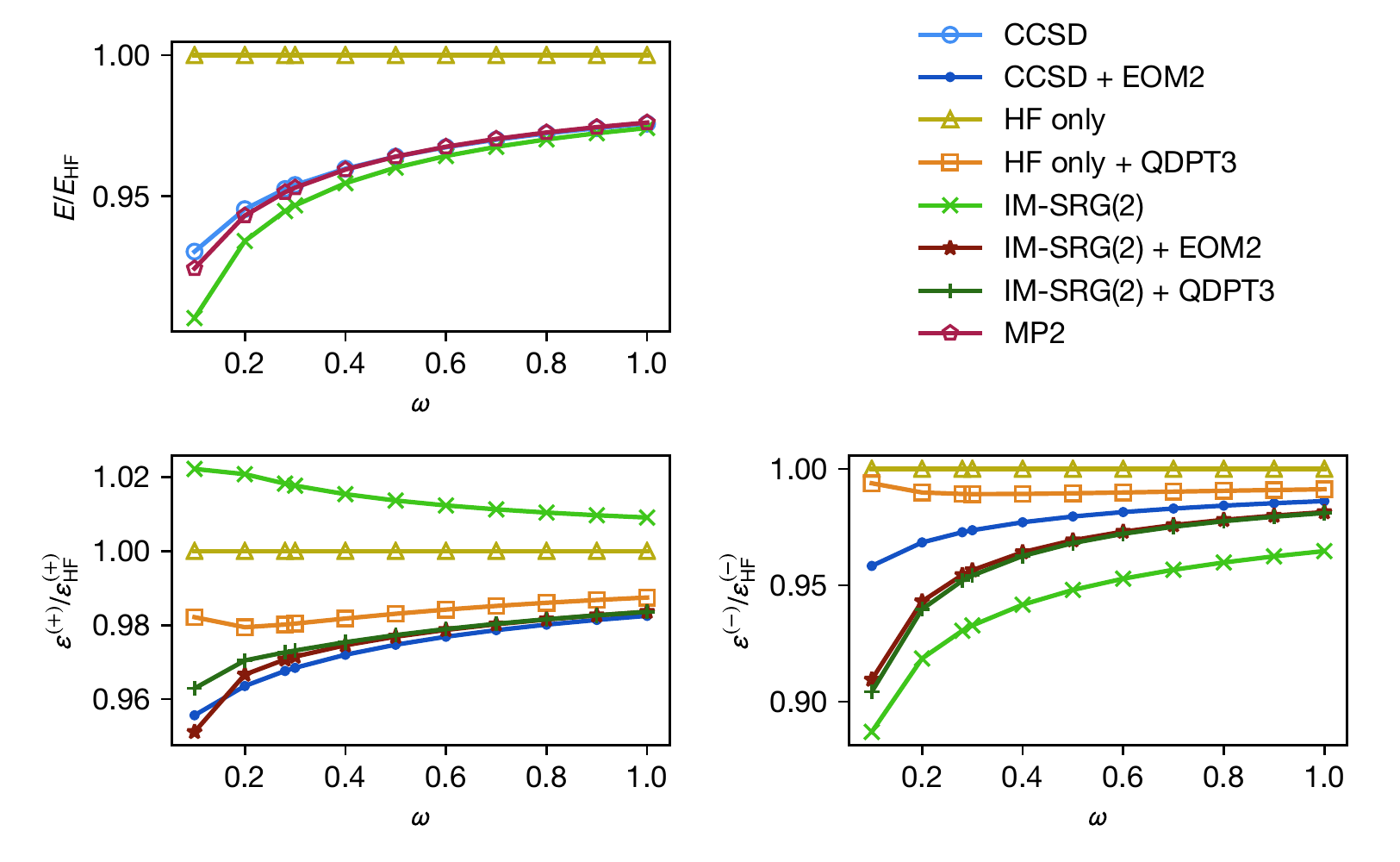}
  \caption{(Color online) The behavior of ground state, addition, and removal energies as a function of the oscillator frequency $\omega$, with $K = 10$ shells in the basis.  The energy is normalized with respect to the HF values to magnify the differences.  Lower frequency leads to stronger correlations and thereby a more difficult problem.}
  \label{fig:by-freq-10-6-normal}
\end{figure}

As before, we do not include results from ``HF only'' in these plots as they are significantly further from the rest.  Analogously, we also exclude results from pure IM-SRG (i.e.\ without QDPT nor EOM) or pure CCSD, as QDPT or EOM both add significant contributions to addition and removal energies.  Some HF only and pure IM-SRG results can be seen in Fig.\ \ref{fig:by-freq-10-6-normal}.

There is strong agreement between IM-SRG(2) + QDPT3 and IM-SRG(2) + EOM2 in many cases, and slightly weaker agreement between the IM-SRG and CCSD families.  This suggests that the EOM2 corrections are largely accounted for by the inexpensive QDPT3 method.  However, in some cases, most notably with few particles and high correlations (low frequency), the IM-SRG(2) + QDPT3 result differs significantly from both IM-SRG(2) + EOM2 and CCSD + EOM2.

\subsection{Rate of convergence}

\begin{figure}
  \centering
  \includegraphics{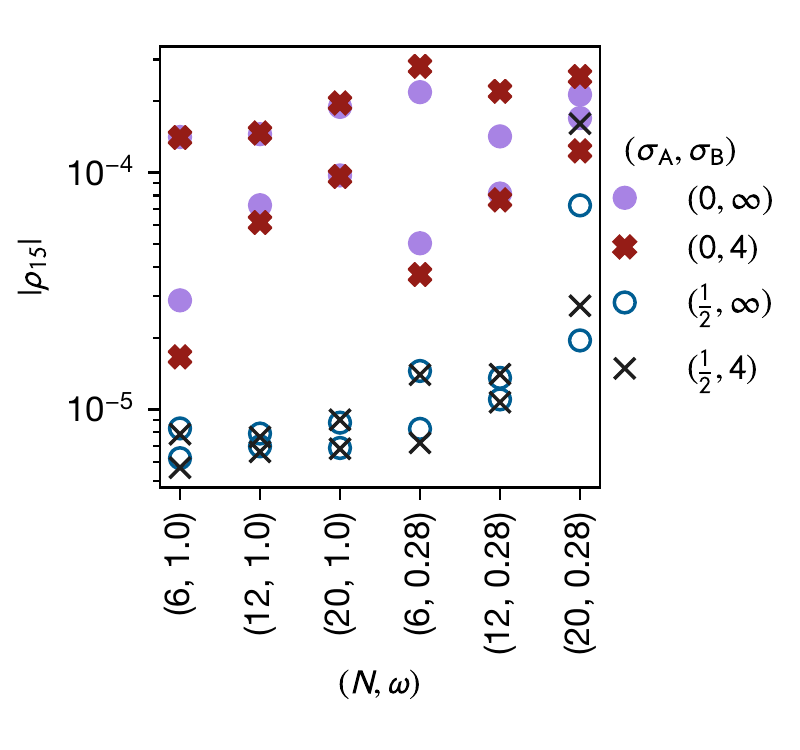}
  \caption{(Color online) The impact of the interaction on convergence of addition and removal energies using IM-SRG(2) + QDPT3.  For clarity, the plot does not distinguish between addition and removal energies.  The horizontal axis shows the system parameters, where $N$ is the number of particles and $\omega$ is the oscillator frequency.  The vertical axis shows $|\rho_{15}|$ (``relative slope''), which estimates the rate of convergence at 15 total shells.  The lower the value of $|\rho_{15}|$, the faster the convergence.  The data points are categorized by the interactions.  The trends suggest that the singular short-range part of the interaction has a much stronger impact on the convergence than the long-range tail.}
  \label{fig:rel-slopes}
\end{figure}

To analyze the rate of convergence more quantitatively, we define $\rho_K$ as the relative backward difference of the energy (\textit{relative slope}):
\begin{align*}
  \rho_K = \frac{\varepsilon_K - \varepsilon_{(K - 1)}}{\varepsilon_K}.
\end{align*}
The denominator allows the quantity to be meaningfully compared between different systems.  We expect this quantity to become increasingly small as the calculations converge towards the complete basis set limit.

In Fig.\ \ref{fig:rel-slopes}, we plot the $\rho_{15}$ for IM-SRG(2) + QDPT3.  The many-body methods were tested against a \emph{modified} Coulomb-like interaction, parametrized by two lengths $\sigma_{\mathrm{A}}$ and $\sigma_{\mathrm{B}}$ that characterize the range of the interaction:
\begin{align}
  V_{\sigma_{\mathrm{A}}, \sigma_{\mathrm{B}}}(r) \equiv \frac{(1 + c)^{1 - 1/c}}{c} \left(1 - \E^{-r^2 / (2 \sigma_{\mathrm{A}}^2)}\right) \E^{-r^2 / (2 \sigma_{\mathrm{B}}^2)} \frac{1}{r},
\end{align}
where $c \equiv \sqrt{\sigma_{\mathrm{B}} / \sigma_{\mathrm{A}}}$.  The coefficient is chosen to ensure the peak of the envelope remains at unity.  With $(\sigma_{\mathrm{A}}, \sigma_{\mathrm{B}}) = (0, \infty)$ one recovers the original Coulomb interaction.  By increasing $\sigma_{\mathrm{A}}$ one can truncate the short-range part of the interaction, and analogously by increasing $\sigma_{\mathrm{B}}$ one can truncate the long-range part of the interaction.  For our numerical experiments we considered the following four combinations of $(\sigma_{\mathrm{A}}, \sigma_{\mathrm{B}})$: $(0, \infty)$, $(\frac{1}{2}, \infty)$, $(0, 4)$, $(\frac{1}{2}, 4)$.

Reducing the short-range part of the interaction appears improves the rate of convergence substantially.  Many of the cases have reached the precision of the ODE solver ($10^{-5}$ to $10^{-6}$).  In contrast, eliminating the long-range part of the interaction had very little effect.  This suggests that the main cause of the slow convergence lies in the highly repulsive, short-ranged part of the interaction, which leads to the presence of nondifferentiable cusps (the so-called \textit{Coulomb cusps}) in the exact wave functions that are difficult to reproduce exactly using linear combinations of the smooth harmonic oscillator wave functions.

The convergence is negatively impacted at lower frequencies and, to a lesser extent, by the increased number of particles.  Both are expected: lower frequencies increase the correlation in the system, while higher number of particles naturally require more shells to converge.

In general, there does not appear to be any difference between the convergence behavior of addition energies as compared to that of removal energies.

\subsection{Extrapolation}
\label{subsec:extrapolation}

\begin{table}
  \centering
  \caption{Extrapolated ground state energies for quantum dots with fit uncertainties, computed from the approximate Hessian in the Levenberg--Marquardt fitting algorithm.  These uncertainties also determine the number of significant figures presented.  Extrapolations are done using 5-point fits where the number of shells $K$ ranges between $K_{\text{stop}} - 4$ and $K_{\text{stop}}$ (inclusive).  The abbreviation ``n.c.'' stands for ``no convergence'': these are extrapolations where, out of the 5 points, at least one of them was unavailable because IM-SRG(2) or CCSD either diverged or converged extremely slowly.}
  \label{tab:ground-extrapolated}
  \begin{tabular}{S[table-format=2.0]SS[table-format=2.0]S[table-format=4.7]S[table-format=4.7]S[table-format=4.7]}%
\hline\hline
{$N$} & {$\omega$} & {$K_{\text{stop}}$} & {MP2} & {IM-SRG(2)} & {CCSD} \\
\hline
6 & 0.1 & 14 & 3.5108(4) & 3.4963(5) & 3.58183(2) \\
6 & 0.28 & 14 & 7.5608(3) & 7.56971(2) & 7.62812(7) \\
6 & 1.0 & 14 & 20.12998(5) & 20.1481(3) & 20.1791(3) \\
\hline
12 & 0.1 & 16 & 12.198(7) & 12.2217(2) & 12.3575(4) \\
12 & 0.28 & 16 & 25.548(1) & 25.6146(1) & 25.7190(2) \\
12 & 1.0 & 16 & 65.627(2) & 65.6970(7) & 65.7579(8) \\
\hline
20 & 0.1 & 16 & 29.87(5) & 29.950(1) & 30.13(2) \\
20 & 0.28 & 16 & 61.88(1) & 61.946(3) & 62.114(5) \\
20 & 1.0 & 16 & 155.758(3) & 155.912(4) & 156.010(3) \\
\hline
30 & 0.1 & 16 & 59.8(2) & {n.c.} & 60.3(2) \\
30 & 0.28 & 16 & 123.95(8) & 124.00(4) & 124.26(5) \\
30 & 1.0 & 16 & 308.80(5) & 308.85(3) & 309.00(3) \\
\hline
42 & 0.1 & 20 & 106.3(4) & 107.0(1) & 107.1(4) \\
42 & 0.28 & 20 & 219.6(2) & 219.89(8) & 220.2(1) \\
42 & 1.0 & 20 & 542.686(9) & 543.074(3) & 543.276(4) \\
\hline
56 & 0.1 & 20 & 172.9(6) & {n.c.} & 174(1) \\
56 & 0.28 & 20 & 357.3(6) & {n.c.} & 358.1(5) \\
56 & 1.0 & 20 & 879.86(8) & 880.07(7) & 880.33(7) \\
\hline\hline
\end{tabular}
\end{table}

\begin{table}
  \centering
  \caption{Extrapolated addition energies for quantum dots with fit uncertainties.  The abbreviation ``n.c.'' has the same meaning as in Table \ref{tab:ground-extrapolated}.  The abbreviation``n.f.'' stands for ``no fit'': this particular extrapolation resulted in unphysical parameters ($\beta \le 0$).  See Table \ref{tab:ground-extrapolated} for other details.}
  \label{tab:add-extrapolated}
  \begin{tabular}{S[table-format=2.0]SS[table-format=2.0]S[table-format=3.8]S[table-format=3.8]S[table-format=3.8]}%
\hline\hline
{$N$} & {$\omega$} & {$K_{\text{stop}}$} & {IM-SRG(2)} & {IMSRG(2)} & {CCSD} \\
{} & {} & {} & {+QDPT3} & {+EOM} & {+EOM} \\
\hline
6 & 0.1 & 14 & 1.206(2) & 1.18095(5) & 1.18581(2) \\
6 & 0.28 & 14 & 2.63(8) & 2.49039(2) & 2.48213(4) \\
6 & 1.0 & 14 & 6.4536(1) & 6.4491(7) & 6.440747(1) \\
\hline
12 & 0.1 & 16 & {{n.f.}} & 1.909274(1) & 1.90139(2) \\
12 & 0.28 & 16 & 3.925(5) & 3.9339(3) & 3.918520(9) \\
12 & 1.0 & 16 & 9.9235(2) & 9.9235(5) & 9.9070(1) \\
\hline
20 & 0.1 & 16 & 2.708(4) & 2.705(3) & 2.682(2) \\
20 & 0.28 & 16 & 5.53915(1) & 5.5405(3) & 5.52180(7) \\
20 & 1.0 & 16 & 13.7759(8) & 13.779(1) & 13.760(2) \\
\hline
30 & 0.1 & 16 & {n.c.} & {n.c.} & 3.40(2) \\
30 & 0.28 & 16 & 7.18(3) & 7.18(5) & 7.16(4) \\
30 & 1.0 & 16 & 17.897(4) & 17.902(6) & 17.880(6) \\
\hline
42 & 0.1 & 20 & 4.19(6) & 4.28(4) & 4.33(2) \\
42 & 0.28 & 20 & 9.068(6) & 9.08(1) & 9.05(1) \\
42 & 1.0 & 20 & 22.2943(7) & 22.30147(1) & 22.2768(9) \\
\hline
56 & 0.1 & 20 & {n.c.} & {n.c.} & 3(3) \\
56 & 0.28 & 20 & {n.c.} & {n.c.} & 10.7(3) \\
56 & 1.0 & 20 & 26.86(4) & 26.87(4) & 26.84(4) \\
\hline\hline
\end{tabular}
\end{table}

\begin{table}
  \centering
  \caption{Extrapolated removal energies for quantum dots with fit uncertainties.  See Table \ref{tab:add-extrapolated} for details.}
  \label{tab:rm-extrapolated}
  \begin{tabular}{S[table-format=2.0]SS[table-format=2.0]S[table-format=3.8]S[table-format=3.8]S[table-format=3.8]}%
\hline\hline
{$N$} & {$\omega$} & {$K_{\text{stop}}$} & {IM-SRG(2)} & {IMSRG(2)} & {CCSD} \\
{} & {} & {} & {+QDPT3} & {+EOM} & {+EOM} \\
\hline
6 & 0.1 & 14 & 0.9509(2) & 0.9561(4) & 1.004943(8) \\
6 & 0.28 & 14 & 2.03396(1) & 2.0387(2) & 2.07620(2) \\
6 & 1.0 & 14 & 5.18889(8) & 5.186(3) & 5.2154(1) \\
\hline
12 & 0.1 & 16 & 1.69624(8) & 1.70181(6) & 1.75031(7) \\
12 & 0.28 & 16 & 3.532236(5) & 3.53512(9) & 3.57527(1) \\
12 & 1.0 & 16 & 8.8039(4) & 8.80390(1) & 8.8331(2) \\
\hline
20 & 0.1 & 16 & 2.5112(6) & 2.5163(8) & 2.55(1) \\
20 & 0.28 & 16 & 5.163(1) & 5.165(1) & 5.208(3) \\
20 & 1.0 & 16 & 12.7122(4) & 12.7101(5) & 12.7442(2) \\
\hline
30 & 0.1 & 16 & {n.c.} & {n.c.} & 3.35(6) \\
30 & 0.28 & 16 & 6.88(2) & 6.88(2) & 6.94(3) \\
30 & 1.0 & 16 & 16.925(2) & 16.923(2) & 16.963(2) \\
\hline
42 & 0.1 & 20 & 4.04(8) & 4.06(7) & 4.1(2) \\
42 & 0.28 & 20 & 8.73(3) & 8.73(3) & 8.76(6) \\
42 & 1.0 & 20 & 21.338(1) & 21.335(1) & 21.378(2) \\
\hline
56 & 0.1 & 20 & {n.c.} & {n.c.} & 5.3(1) \\
56 & 0.28 & 20 & {n.c.} & {n.c.} & 10.75(9) \\
56 & 1.0 & 20 & 26.008(9) & 26.004(8) & 26.050(9) \\
\hline\hline
\end{tabular}
\end{table}

\begin{figure}
  \centering
  \includegraphics{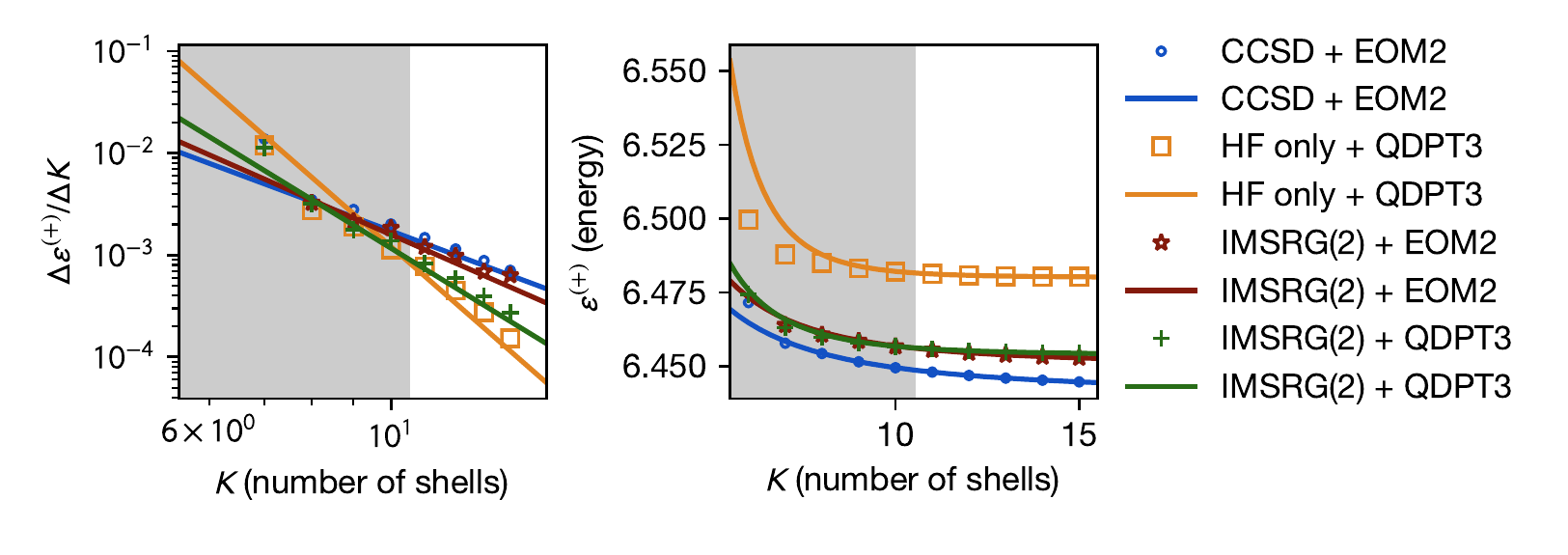}
  \caption{(Color online) A five-point fit of the addition energies of the $(N, \omega) = (6, 1.0)$ system with $K_{\text{stop}} = 15$.  The grey shaded region contains the masked data points, which are ignored by the fitting procedure.  The left figure plots the central difference of the addition energies $\varepsilon^{(+)}$ with respect to the number of shells $K$.  On such a plot, the power law model should appear as a straight line.  The fits are optimized using the procedure described in Section \ref{subsec:extrapolation}.  Note that the lines do not evenly pass through the points in the left plot as the fitting weights are tuned for the energy on a linear scale, not the energy differences on a logarithmic scale.}
  \label{fig:by-fit-2-1p0-add}
\end{figure}

To reduce errors from the basis set truncation, one can either use explicitly correlated R12/F12 methods that account for the correct cusp behavior in many-electron wave functions\cite{Kutzelnigg1985,KLOPPER198717,explicitcorrelationreview}, or one can use basis extrapolation techniques.  In the present work, we focus on the latter.  As derived by Kvaal \cite{PhysRevB.80.045321,Kvaal2007}, the asymptotic convergence of quantum dot observables in a finite harmonic oscillator basis can be approximately described by a power law model:
\begin{align*}
  \Delta E \propto K^{-\beta},
\end{align*}
where $\Delta E$ is the difference between the finite-basis result and the infinite-basis result, $K$ is the number of shells in the single-particle basis, and $\beta$ is some positive real exponent.  The \textit{smoothness} of the exact wave function determines the rate of the convergence: the more times the exact wave function can be differentiated, the higher the exponent $\beta$.

We note that this model was derived under the assumption that all correlations are included in the calculation (i.e.\ FCI), thus we are making an assumption that our selection of methods approximately obey the same behavior.  The validity of this assumption will be assessed at the end of this section.

In general, the exponent $\beta$ cannot be determined \textit{a priori}, thus we will empirically compute $\beta$ by fitting the following model through our data:
\begin{align} \label{eq:power_law_model}
  E = \alpha K^{-\beta} + \gamma.
\end{align}
As a nonlinear curve fit, it can be quite sensitive to the initial parameters.  Therefore, good guesses of the parameters are necessary to obtain a sensible result.  For this, we first fit a linear model of $\log |\partial E / \partial K|$ against $\log K$:
\begin{align*}
  \log \left|\frac{\partial E}{\partial K}\right| = - (\beta + 1) \log K + \log|\alpha \beta|.
\end{align*}
This is useful because linear fits are very robust and will often converge even if the initial parameters are far from their final values.  It also provides a means to visually assess the quality of the fit.  The derivative is approximated using the central difference:
\begin{align*}
  \frac{\partial E}{\partial K} \approx E\left(K + \frac{1}{2}\right) - E\left(K - \frac{1}{2}\right).
\end{align*}
The process of numerically calculating the derivative can amplify the noise in the data and distorts the weights of the data points.  Moreover, it does not provide a means to compute $\gamma$, the extrapolated energy.  Thus a second accurate nonlinear curve fit is necessary.

The parameters $\alpha$ and $\beta$ are extracted from the linear fit and used as inputs for a power-law fit of $E$ against $K$.  It is necessary to estimate the infinite-basis energy $\gamma$ as well, which is done by fitting Eq.\ \eqref{eq:power_law_model} while the parameters $\alpha$ and $\beta$ are fixed to the initial guesses.  The fixing ensures that the fit is still linear in nature and thus highly likely to converge.  Afterward, we do a final fit with all three parameters free to vary.  All fits are done using the traditional Levenberg--Marquardt (LM) optimization algorithm \cite{10.2307/43633451,doi:10.1137/0111030} as implemented in \texttt{Minpack} \cite{More1978,More:126569}, with equal weighting of all data points.

There is still one additional tuning knob for this model that is not explicitly part of Eq.\ \eqref{eq:power_law_model}: the range of data points taken into consideration (\textit{fit range}).  Since the model describes the \emph{asymptotic} behavior, we do not expect the fit to produce good results when the energy is still very far from convergence.  To account for this, we only fit the last few data points within some chosen range.  If the range is too large, then the non-asymptotic behavior would perturb the result too much, whereas if the range is too small, there would be more noise and less confidence in whether the trend is legitimate rather than accidental.  Empirically, we chose to fit the last 5 points of our available data.  The results are shown in Tables \ref{tab:ground-extrapolated}, \ref{tab:add-extrapolated}, and \ref{tab:rm-extrapolated}.  A specific example of the fit is shown in Fig.\ \ref{fig:by-fit-2-1p0-add}

The LM fitting procedure also computes uncertainties for the parameters from an approximate Hessian of the model function.  It is therefore tempting to use the uncertainty of the fit to quantify the uncertainty of the extrapolated energy.  We certainly would not expect this to account for the error due to the operator truncation, but how accurately does it quantify the discrepancy of our extrapolated result from the true infinite-basis energy?

We investigated this idea by performing a fit over \emph{all possible} 5-point fit ranges $[K_{\text{stop}} - 4, K_{\text{stop}}]$.  By comparing the extrapolated results at varying values of $K_{\text{stop}}$ with the extrapolated result at the highest possible $K_{\text{stop}}$ and treating the latter as the ``true'' infinite-basis result, we can statistically assess whether the fit uncertainties are a good measure of the discrepancy from the true infinite-basis result.  Our results show a somewhat bimodal distribution: when the relative fit uncertainty is higher than $10^{-3.5}$, the fit uncertainty quantifies the discrepancy well; otherwise, the fit uncertainty underestimates the discrepancy by a factor of 10 or less.

Unlike the other methods, HF energies are somewhat unusual in that they generally do not conform to the power-law model.  In fact, the plots indicate an \emph{exponential} convergence with respect to the number of shells, which has also been observed in molecular systems.\cite{HALKIER1999437}  We surmise that HF is insensitive to the Coulomb cusp.

Nonetheless, despite the poor fits that often arise, the extrapolated energies are often quite good for HF.  This is likely due to its rapid convergence, which leaves very little degree of freedom even for a poorly chosen model.  Moreover, we found that the fit uncertainties of the energy are fairly good measures of the true discrepancy.

Not all fits yield a positive value of $\beta$ for addition and removal energies, which suggests that the data points do not converge, or require a very high number of shells to converge.  This affects exclusively IM-SRG(2) + QDPT3 for systems with few particles and low frequencies, indicating that perturbation theory is inadequate for such systems.

\section{Conclusions}
\label{sec:conclusions}

We have demonstrated calculations of ground state, addition, and removal energies of two-dimensional circular quantum dots using a variety of many-body methods: ranging from the basic HF method, to more sophisticated combinations of IM-SRG, CC, QDPT and/or EOM.  Many closed-shell quantum dot systems have been explored, ranging from 2 to 56 particles and frequencies between 0.1 and 1.0.  All such results show good agreement with one another.

We note that the HF + IM-SRG + QDPT combination provides a reasonable moderate-cost approach to the calculation of addition and removal energies for many systems in comparison to the somewhat more expensive and complicated EOM calculations.  Both IM-SRG and CC are reasonably accurate compared to the near-exact but factorial-cost FCI method, allowing exploration of much higher number of particles than would otherwise be possible.  EOM-IM-SRG does have more flexibility over perturbative approaches: in particular, it could be more readily used to construct excited states \cite{PhysRevC.95.044304}, which is more difficult for methods such as DMC.

There are several directions in which the calculations may be improved.  One can attempt to improve the IM-SRG approximation by incorporating some of the missing higher-body terms in the commutator.  This would also provide some insight into the rate of convergence with respect to the operator truncation, providing a sense of how large the truncation error is.  While a full 3-body treatment of IM-SRG would be extremely costly, it is possible implicitly track for a portion of the induced 3-body forces by computing certain diagrams that are either lower cost or could be approximated at lower cost \cite{Hergert2016165}.

The application of IM-SRG eliminates a large number of the QDPT diagrams of Fig.\ \ref{fig:diagrams-sfe}, which is beneficial as it increases the efficiency of the QDPT calculations.  Thus it may be more feasible to perform higher orders of QDPT on IM-SRG evolved Hamiltonians.  Moreover, the remaining diagrams at third order can be eliminated through infinite resummation techniques, which would incorporate higher order terms and therefore further increase the accuracy of the result.

We note that this calculation was done entirely using the traditional approach of using a high-order ODE solver to solve the flow equation.  A new technique developed by Morris et al.\ \cite{PhysRevC.92.034331} uses an alternative approach based on the Magnus expansion that obviates the need for a high-order ODE solver, leading to much more efficient computations and also allowing operators of other observables to be evolved at lower cost, which presents a significant advantage over CC methods.  Implementing this approach would allows us to study the accuracy and convergence of other possibly more sensitive observables.

The IM-SRG method can be extended to support multiple reference states (multi-reference IM-SRG or simply MR-IM-SRG) \cite{PhysRevLett.110.242501,PhysRevC.90.041302} through the generalized normal ordering formalisms \cite{doi:10.1063/1.474405}, which opens the possibility of calculating quantum systems that are far from the magic numbers (open-shell systems).  This is a territory that few many-body methods can tackle and is one of the major strengths of the IM-SRG approach.

By transforming of the operator-SRG rather than the wave function, IM-SRG is more amenable to the construction of softened \textit{effective interactions} than CC \cite{Hergert2016165}.  Such interactions can be used to lower the cost of other methods such as full configuration interaction theory.

Implementation-wise, our current code constructs the two-particle states as simple Slater determinants of their one-particle states.  This approach is often referred to as \textit{m-scheme} in nuclear physics.  It is straightforward to implement and works well for general systems, but it fails to exploit all the symmetries in the quantum dot system: not only is spin projection $\hat S_z$ conserved, the Casimir operator $\hat S^2$ is as well.  A more efficient approach is to couple the spins of the single-particle states to form two-particle states with good $\hat S^2$, an approach analogous to \textit{j-scheme} in nuclear physics.  A combination of this with the Wigner--Eckart theorem could reduce the computational cost significantly, albeit at the cost of increased implementation complexity.

We hope to apply these theoretical and technical enhancements not only to the study of quantum dots, but to studies of nuclear and atomic systems, opening up an even greater range of applications.

\begin{acknowledgments}
  This work was supported by the National Science Foundation Grant No.\ PHY-1404159 (Michigan State University).  Discussions with Simen Kvaal are highly appreciated.
\end{acknowledgments}

\bibliography{paper}

\begin{thebibliography}{82}%
\makeatletter
\providecommand \@ifxundefined [1]{%
 \@ifx{#1\undefined}
}%
\providecommand \@ifnum [1]{%
 \ifnum #1\expandafter \@firstoftwo
 \else \expandafter \@secondoftwo
 \fi
}%
\providecommand \@ifx [1]{%
 \ifx #1\expandafter \@firstoftwo
 \else \expandafter \@secondoftwo
 \fi
}%
\providecommand \natexlab [1]{#1}%
\providecommand \enquote  [1]{``#1''}%
\providecommand \bibnamefont  [1]{#1}%
\providecommand \bibfnamefont [1]{#1}%
\providecommand \citenamefont [1]{#1}%
\providecommand \href@noop [0]{\@secondoftwo}%
\providecommand \href [0]{\begingroup \@sanitize@url \@href}%
\providecommand \@href[1]{\@@startlink{#1}\@@href}%
\providecommand \@@href[1]{\endgroup#1\@@endlink}%
\providecommand \@sanitize@url [0]{\catcode `\\12\catcode `\$12\catcode
  `\&12\catcode `\#12\catcode `\^12\catcode `\_12\catcode `\%12\relax}%
\providecommand \@@startlink[1]{}%
\providecommand \@@endlink[0]{}%
\providecommand \url  [0]{\begingroup\@sanitize@url \@url }%
\providecommand \@url [1]{\endgroup\@href {#1}{\urlprefix }}%
\providecommand \urlprefix  [0]{URL }%
\providecommand \Eprint [0]{\href }%
\providecommand \doibase [0]{http://dx.doi.org/}%
\providecommand \selectlanguage [0]{\@gobble}%
\providecommand \bibinfo  [0]{\@secondoftwo}%
\providecommand \bibfield  [0]{\@secondoftwo}%
\providecommand \translation [1]{[#1]}%
\providecommand \BibitemOpen [0]{}%
\providecommand \bibitemStop [0]{}%
\providecommand \bibitemNoStop [0]{.\EOS\space}%
\providecommand \EOS [0]{\spacefactor3000\relax}%
\providecommand \BibitemShut  [1]{\csname bibitem#1\endcsname}%
\let\auto@bib@innerbib\@empty
\bibitem [{\citenamefont {Reimann}\ and\ \citenamefont
  {Manninen}(2002)}]{reimann2002}%
  \BibitemOpen
  \bibfield  {author} {\bibinfo {author} {\bibfnamefont {S.~M.}\ \bibnamefont
  {Reimann}}\ and\ \bibinfo {author} {\bibfnamefont {M.}~\bibnamefont
  {Manninen}},\ }\href@noop {} {\bibfield  {journal} {\bibinfo  {journal} {Rev.
  Mod. Phys.}\ }\textbf {\bibinfo {volume} {74}},\ \bibinfo {pages} {1238}
  (\bibinfo {year} {2002})}\BibitemShut {NoStop}%
\bibitem [{\citenamefont {Engel}\ \emph {et~al.}(2004)\citenamefont {Engel},
  \citenamefont {Golovach}, \citenamefont {Loss}, \citenamefont {Vandersypen},
  \citenamefont {Elzerman}, \citenamefont {Hanson},\ and\ \citenamefont
  {Kouwenhoven}}]{engel1993}%
  \BibitemOpen
  \bibfield  {author} {\bibinfo {author} {\bibfnamefont {H.-A.}\ \bibnamefont
  {Engel}}, \bibinfo {author} {\bibfnamefont {V.~N.}\ \bibnamefont {Golovach}},
  \bibinfo {author} {\bibfnamefont {D.}~\bibnamefont {Loss}}, \bibinfo {author}
  {\bibfnamefont {L.~M.~K.}\ \bibnamefont {Vandersypen}}, \bibinfo {author}
  {\bibfnamefont {J.~M.}\ \bibnamefont {Elzerman}}, \bibinfo {author}
  {\bibfnamefont {R.}~\bibnamefont {Hanson}}, \ and\ \bibinfo {author}
  {\bibfnamefont {L.~P.}\ \bibnamefont {Kouwenhoven}},\ }\href {\doibase
  10.1103/PhysRevLett.93.106804} {\bibfield  {journal} {\bibinfo  {journal}
  {Phys. Rev. Lett.}\ }\textbf {\bibinfo {volume} {93}},\ \bibinfo {pages}
  {106804} (\bibinfo {year} {2004})}\BibitemShut {NoStop}%
\bibitem [{\citenamefont {Birman}, \citenamefont {Nazmitdinov},\ and\
  \citenamefont {Yukalov}(2013)}]{BIRMAN20131}%
  \BibitemOpen
  \bibfield  {author} {\bibinfo {author} {\bibfnamefont {J.}~\bibnamefont
  {Birman}}, \bibinfo {author} {\bibfnamefont {R.}~\bibnamefont {Nazmitdinov}},
  \ and\ \bibinfo {author} {\bibfnamefont {V.}~\bibnamefont {Yukalov}},\ }\href
  {\doibase 10.1016/j.physrep.2012.11.005} {\bibfield  {journal} {\bibinfo
  {journal} {Phys. Rep.}\ }\textbf {\bibinfo {volume} {526}},\ \bibinfo {pages}
  {1 } (\bibinfo {year} {2013})},\ \bibinfo {note} {effects of symmetry
  breaking in finite quantum systems}\BibitemShut {NoStop}%
\bibitem [{\citenamefont {Tarucha}\ \emph {et~al.}(1996)\citenamefont
  {Tarucha}, \citenamefont {Austing}, \citenamefont {Honda}, \citenamefont
  {van~der Hage},\ and\ \citenamefont {Kouwenhoven}}]{tarucha1996}%
  \BibitemOpen
  \bibfield  {author} {\bibinfo {author} {\bibfnamefont {S.}~\bibnamefont
  {Tarucha}}, \bibinfo {author} {\bibfnamefont {D.~G.}\ \bibnamefont
  {Austing}}, \bibinfo {author} {\bibfnamefont {T.}~\bibnamefont {Honda}},
  \bibinfo {author} {\bibfnamefont {R.~J.}\ \bibnamefont {van~der Hage}}, \
  and\ \bibinfo {author} {\bibfnamefont {L.~P.}\ \bibnamefont {Kouwenhoven}},\
  }\href {\doibase 10.1103/PhysRevLett.77.3613} {\bibfield  {journal} {\bibinfo
   {journal} {Phys. Rev. Lett.}\ }\textbf {\bibinfo {volume} {77}},\ \bibinfo
  {pages} {3613} (\bibinfo {year} {1996})}\BibitemShut {NoStop}%
\bibitem [{\citenamefont {Bogner}\ \emph {et~al.}(2011)\citenamefont {Bogner},
  \citenamefont {Furnstahl}, \citenamefont {Hergert}, \citenamefont
  {Kortelainen}, \citenamefont {Maris}, \citenamefont {Stoitsov},\ and\
  \citenamefont {Vary}}]{PhysRevC.84.044306}%
  \BibitemOpen
  \bibfield  {author} {\bibinfo {author} {\bibfnamefont {S.~K.}\ \bibnamefont
  {Bogner}}, \bibinfo {author} {\bibfnamefont {R.~J.}\ \bibnamefont
  {Furnstahl}}, \bibinfo {author} {\bibfnamefont {H.}~\bibnamefont {Hergert}},
  \bibinfo {author} {\bibfnamefont {M.}~\bibnamefont {Kortelainen}}, \bibinfo
  {author} {\bibfnamefont {P.}~\bibnamefont {Maris}}, \bibinfo {author}
  {\bibfnamefont {M.}~\bibnamefont {Stoitsov}}, \ and\ \bibinfo {author}
  {\bibfnamefont {J.~P.}\ \bibnamefont {Vary}},\ }\href {\doibase
  10.1103/PhysRevC.84.044306} {\bibfield  {journal} {\bibinfo  {journal} {Phys.
  Rev. C}\ }\textbf {\bibinfo {volume} {84}},\ \bibinfo {pages} {044306}
  (\bibinfo {year} {2011})}\BibitemShut {NoStop}%
\bibitem [{\citenamefont {Jenks}\ and\ \citenamefont
  {Gilmore}(2010)}]{jenks:013111}%
  \BibitemOpen
  \bibfield  {author} {\bibinfo {author} {\bibfnamefont {S.}~\bibnamefont
  {Jenks}}\ and\ \bibinfo {author} {\bibfnamefont {R.}~\bibnamefont
  {Gilmore}},\ }\href@noop {} {\bibfield  {journal} {\bibinfo  {journal} {J.
  Renew. Sustain. Ener.}\ }\textbf {\bibinfo {volume} {2}},\ \bibinfo {pages}
  {013111} (\bibinfo {year} {2010})}\BibitemShut {NoStop}%
\bibitem [{\citenamefont {Nozik}\ \emph {et~al.}(2010)\citenamefont {Nozik},
  \citenamefont {Beard}, \citenamefont {Luther}, \citenamefont {Law},
  \citenamefont {Ellingson},\ and\ \citenamefont
  {Johnson}}]{doi:10.1021/cr900289f}%
  \BibitemOpen
  \bibfield  {author} {\bibinfo {author} {\bibfnamefont {A.~J.}\ \bibnamefont
  {Nozik}}, \bibinfo {author} {\bibfnamefont {M.~C.}\ \bibnamefont {Beard}},
  \bibinfo {author} {\bibfnamefont {J.~M.}\ \bibnamefont {Luther}}, \bibinfo
  {author} {\bibfnamefont {M.}~\bibnamefont {Law}}, \bibinfo {author}
  {\bibfnamefont {R.~J.}\ \bibnamefont {Ellingson}}, \ and\ \bibinfo {author}
  {\bibfnamefont {J.~C.}\ \bibnamefont {Johnson}},\ }\href {\doibase
  10.1021/cr900289f} {\bibfield  {journal} {\bibinfo  {journal} {Chem. Rev.}\
  }\textbf {\bibinfo {volume} {110}},\ \bibinfo {pages} {6873} (\bibinfo {year}
  {2010})}\BibitemShut {NoStop}%
\bibitem [{\citenamefont {Strauf}\ \emph {et~al.}(2006)\citenamefont {Strauf},
  \citenamefont {Hennessy}, \citenamefont {Rakher}, \citenamefont {Choi},
  \citenamefont {Badolato}, \citenamefont {Andreani}, \citenamefont {Hu},
  \citenamefont {Petroff},\ and\ \citenamefont {Bouwmeester}}]{strauf2010}%
  \BibitemOpen
  \bibfield  {author} {\bibinfo {author} {\bibfnamefont {S.}~\bibnamefont
  {Strauf}}, \bibinfo {author} {\bibfnamefont {K.}~\bibnamefont {Hennessy}},
  \bibinfo {author} {\bibfnamefont {M.}~\bibnamefont {Rakher}}, \bibinfo
  {author} {\bibfnamefont {Y.-S.}\ \bibnamefont {Choi}}, \bibinfo {author}
  {\bibfnamefont {A.}~\bibnamefont {Badolato}}, \bibinfo {author}
  {\bibfnamefont {L.~C.}\ \bibnamefont {Andreani}}, \bibinfo {author}
  {\bibfnamefont {E.~L.}\ \bibnamefont {Hu}}, \bibinfo {author} {\bibfnamefont
  {P.}~\bibnamefont {Petroff}}, \ and\ \bibinfo {author} {\bibfnamefont
  {D.}~\bibnamefont {Bouwmeester}},\ }\href {\doibase
  10.1103/physrevlett.96.127404} {\bibfield  {journal} {\bibinfo  {journal}
  {Phys. Rev. Lett.}\ }\textbf {\bibinfo {volume} {96}},\ \bibinfo {pages}
  {127404} (\bibinfo {year} {2006})}\BibitemShut {NoStop}%
\bibitem [{\citenamefont {Mi}\ \emph {et~al.}(2009)\citenamefont {Mi},
  \citenamefont {Yang}, \citenamefont {Bhattacharya}, \citenamefont {Qin},\
  and\ \citenamefont {Ma}}]{5075760}%
  \BibitemOpen
  \bibfield  {author} {\bibinfo {author} {\bibfnamefont {Z.}~\bibnamefont
  {Mi}}, \bibinfo {author} {\bibfnamefont {J.}~\bibnamefont {Yang}}, \bibinfo
  {author} {\bibfnamefont {P.}~\bibnamefont {Bhattacharya}}, \bibinfo {author}
  {\bibfnamefont {G.}~\bibnamefont {Qin}}, \ and\ \bibinfo {author}
  {\bibfnamefont {Z.}~\bibnamefont {Ma}},\ }\href {\doibase
  10.1109/JPROC.2009.2014780} {\bibfield  {journal} {\bibinfo  {journal} {Proc.
  IEEE}\ }\textbf {\bibinfo {volume} {97}},\ \bibinfo {pages} {1239} (\bibinfo
  {year} {2009})}\BibitemShut {NoStop}%
\bibitem [{\citenamefont {Ben-Ari}(2003)}]{Ben-Ari02042003}%
  \BibitemOpen
  \bibfield  {author} {\bibinfo {author} {\bibfnamefont {E.~T.}\ \bibnamefont
  {Ben-Ari}},\ }\href {\doibase 10.1093/jnci} {\bibfield  {journal} {\bibinfo
  {journal} {J. Natl. Cancer Inst.}\ }\textbf {\bibinfo {volume} {95}},\
  \bibinfo {pages} {502} (\bibinfo {year} {2003})}\BibitemShut {NoStop}%
\bibitem [{\citenamefont {Loss}\ and\ \citenamefont
  {DiVincenzo}(1998)}]{PhysRevA.57.120}%
  \BibitemOpen
  \bibfield  {author} {\bibinfo {author} {\bibfnamefont {D.}~\bibnamefont
  {Loss}}\ and\ \bibinfo {author} {\bibfnamefont {D.~P.}\ \bibnamefont
  {DiVincenzo}},\ }\href {\doibase 10.1103/PhysRevA.57.120} {\bibfield
  {journal} {\bibinfo  {journal} {Phys. Rev. A}\ }\textbf {\bibinfo {volume}
  {57}},\ \bibinfo {pages} {120} (\bibinfo {year} {1998})}\BibitemShut
  {NoStop}%
\bibitem [{\citenamefont {Taut}(1993)}]{PhysRevA.48.3561}%
  \BibitemOpen
  \bibfield  {author} {\bibinfo {author} {\bibfnamefont {M.}~\bibnamefont
  {Taut}},\ }\href {\doibase 10.1103/PhysRevA.48.3561} {\bibfield  {journal}
  {\bibinfo  {journal} {Phys. Rev. A}\ }\textbf {\bibinfo {volume} {48}},\
  \bibinfo {pages} {3561} (\bibinfo {year} {1993})}\BibitemShut {NoStop}%
\bibitem [{\citenamefont {Taut}(1994)}]{10.1088/0305-4470/27/3/040}%
  \BibitemOpen
  \bibfield  {author} {\bibinfo {author} {\bibfnamefont {M.}~\bibnamefont
  {Taut}},\ }\href {\doibase 10.1088/0305-4470/27/3/040} {\bibfield  {journal}
  {\bibinfo  {journal} {J. Phys. A Math. Gen.}\ }\textbf {\bibinfo {volume}
  {27}},\ \bibinfo {pages} {1045} (\bibinfo {year} {1994})}\BibitemShut
  {NoStop}%
\bibitem [{\citenamefont {Hartree}(1928)}]{hartree_1928}%
  \BibitemOpen
  \bibfield  {author} {\bibinfo {author} {\bibfnamefont {D.~R.}\ \bibnamefont
  {Hartree}},\ }\href {\doibase 10.1017/S0305004100011919} {\bibfield
  {journal} {\bibinfo  {journal} {Math. Proc. Cambridge}\ }\textbf {\bibinfo
  {volume} {24}},\ \bibinfo {pages} {89} (\bibinfo {year} {1928})}\BibitemShut
  {NoStop}%
\bibitem [{\citenamefont {Fock}(1930)}]{Fock1930}%
  \BibitemOpen
  \bibfield  {author} {\bibinfo {author} {\bibfnamefont {V.}~\bibnamefont
  {Fock}},\ }\href {\doibase 10.1007/BF01340294} {\bibfield  {journal}
  {\bibinfo  {journal} {Z. Phys.}\ }\textbf {\bibinfo {volume} {61}},\ \bibinfo
  {pages} {126} (\bibinfo {year} {1930})}\BibitemShut {NoStop}%
\bibitem [{\citenamefont {{M{\o}ller}}\ and\ \citenamefont
  {{Plesset}}(1934)}]{1934PhRv...46..618M}%
  \BibitemOpen
  \bibfield  {author} {\bibinfo {author} {\bibfnamefont {C.}~\bibnamefont
  {{M{\o}ller}}}\ and\ \bibinfo {author} {\bibfnamefont {M.~S.}\ \bibnamefont
  {{Plesset}}},\ }\href {\doibase 10.1103/PhysRev.46.618} {\bibfield  {journal}
  {\bibinfo  {journal} {Phys. Rev.}\ }\textbf {\bibinfo {volume} {46}},\
  \bibinfo {pages} {618} (\bibinfo {year} {1934})}\BibitemShut {NoStop}%
\bibitem [{\citenamefont {Hergert}\ \emph {et~al.}(2016)\citenamefont
  {Hergert}, \citenamefont {Bogner}, \citenamefont {Morris}, \citenamefont
  {Schwenk},\ and\ \citenamefont {Tsukiyama}}]{Hergert2016165}%
  \BibitemOpen
  \bibfield  {author} {\bibinfo {author} {\bibfnamefont {H.}~\bibnamefont
  {Hergert}}, \bibinfo {author} {\bibfnamefont {S.}~\bibnamefont {Bogner}},
  \bibinfo {author} {\bibfnamefont {T.}~\bibnamefont {Morris}}, \bibinfo
  {author} {\bibfnamefont {A.}~\bibnamefont {Schwenk}}, \ and\ \bibinfo
  {author} {\bibfnamefont {K.}~\bibnamefont {Tsukiyama}},\ }\href {\doibase
  10.1016/j.physrep.2015.12.007} {\bibfield  {journal} {\bibinfo  {journal}
  {Phys. Rep.}\ }\textbf {\bibinfo {volume} {621}},\ \bibinfo {pages} {165}
  (\bibinfo {year} {2016})}\BibitemShut {NoStop}%
\bibitem [{\citenamefont {Henderson}, \citenamefont {Runge},\ and\
  \citenamefont {Bartlett}(2003)}]{PhysRevB.67.045320}%
  \BibitemOpen
  \bibfield  {author} {\bibinfo {author} {\bibfnamefont {T.~M.}\ \bibnamefont
  {Henderson}}, \bibinfo {author} {\bibfnamefont {K.}~\bibnamefont {Runge}}, \
  and\ \bibinfo {author} {\bibfnamefont {R.~J.}\ \bibnamefont {Bartlett}},\
  }\href {\doibase 10.1103/PhysRevB.67.045320} {\bibfield  {journal} {\bibinfo
  {journal} {Phys. Rev. B}\ }\textbf {\bibinfo {volume} {67}},\ \bibinfo
  {pages} {045320} (\bibinfo {year} {2003})}\BibitemShut {NoStop}%
\bibitem [{\citenamefont {Heidari}\ \emph {et~al.}(2007)\citenamefont
  {Heidari}, \citenamefont {Pal}, \citenamefont {Pujari},\ and\ \citenamefont
  {Kanhere}}]{heidari:114708}%
  \BibitemOpen
  \bibfield  {author} {\bibinfo {author} {\bibfnamefont {I.}~\bibnamefont
  {Heidari}}, \bibinfo {author} {\bibfnamefont {S.}~\bibnamefont {Pal}},
  \bibinfo {author} {\bibfnamefont {B.~S.}\ \bibnamefont {Pujari}}, \ and\
  \bibinfo {author} {\bibfnamefont {D.~G.}\ \bibnamefont {Kanhere}},\ }\href
  {\doibase 10.1063/1.2768523} {\bibfield  {journal} {\bibinfo  {journal} {J.
  Chem. Phys.}\ }\textbf {\bibinfo {volume} {127}},\ \bibinfo {pages} {114708}
  (\bibinfo {year} {2007})}\BibitemShut {NoStop}%
\bibitem [{\citenamefont {{Pedersen Lohne}}\ \emph {et~al.}(2011)\citenamefont
  {{Pedersen Lohne}}, \citenamefont {Hagen}, \citenamefont {Hjorth-Jensen},
  \citenamefont {Kvaal},\ and\ \citenamefont {Pederiva}}]{PhysRevB.84.115302}%
  \BibitemOpen
  \bibfield  {author} {\bibinfo {author} {\bibfnamefont {M.}~\bibnamefont
  {{Pedersen Lohne}}}, \bibinfo {author} {\bibfnamefont {G.}~\bibnamefont
  {Hagen}}, \bibinfo {author} {\bibfnamefont {M.}~\bibnamefont
  {Hjorth-Jensen}}, \bibinfo {author} {\bibfnamefont {S.}~\bibnamefont
  {Kvaal}}, \ and\ \bibinfo {author} {\bibfnamefont {F.}~\bibnamefont
  {Pederiva}},\ }\href@noop {} {\bibfield  {journal} {\bibinfo  {journal}
  {Phys. Rev. B}\ }\textbf {\bibinfo {volume} {84}},\ \bibinfo {pages} {115302}
  (\bibinfo {year} {2011})}\BibitemShut {NoStop}%
\bibitem [{\citenamefont {Lindgren}(1974)}]{0022-3700-7-18-010}%
  \BibitemOpen
  \bibfield  {author} {\bibinfo {author} {\bibfnamefont {I.}~\bibnamefont
  {Lindgren}},\ }\href {\doibase 10.1088/0022-3700/7/18/010} {\bibfield
  {journal} {\bibinfo  {journal} {J. Phys. Pt. B Atom. M. P.}\ }\textbf
  {\bibinfo {volume} {7}},\ \bibinfo {pages} {2441} (\bibinfo {year}
  {1974})}\BibitemShut {NoStop}%
\bibitem [{\citenamefont {Kvasni{\v{c}}ka}(1974)}]{Kvasnicka1974}%
  \BibitemOpen
  \bibfield  {author} {\bibinfo {author} {\bibfnamefont {V.}~\bibnamefont
  {Kvasni{\v{c}}ka}},\ }\href {\doibase 10.1007/BF01587295} {\bibfield
  {journal} {\bibinfo  {journal} {Czech. J. Phys. Sect. B}\ }\textbf {\bibinfo
  {volume} {24}},\ \bibinfo {pages} {605} (\bibinfo {year} {1974})}\BibitemShut
  {NoStop}%
\bibitem [{\citenamefont {Rowe}(1968)}]{RevModPhys.40.153}%
  \BibitemOpen
  \bibfield  {author} {\bibinfo {author} {\bibfnamefont {D.~J.}\ \bibnamefont
  {Rowe}},\ }\href {\doibase 10.1103/RevModPhys.40.153} {\bibfield  {journal}
  {\bibinfo  {journal} {Rev. Mod. Phys.}\ }\textbf {\bibinfo {volume} {40}},\
  \bibinfo {pages} {153} (\bibinfo {year} {1968})}\BibitemShut {NoStop}%
\bibitem [{\citenamefont {Stanton}\ and\ \citenamefont
  {Bartlett}(1993)}]{StantonBartlettEOM}%
  \BibitemOpen
  \bibfield  {author} {\bibinfo {author} {\bibfnamefont {J.~F.}\ \bibnamefont
  {Stanton}}\ and\ \bibinfo {author} {\bibfnamefont {R.~J.}\ \bibnamefont
  {Bartlett}},\ }\href {\doibase 10.1063/1.464746} {\bibfield  {journal}
  {\bibinfo  {journal} {J. Chem. Phys.}\ }\textbf {\bibinfo {volume} {98}},\
  \bibinfo {pages} {7029} (\bibinfo {year} {1993})}\BibitemShut {NoStop}%
\bibitem [{\citenamefont {Emrich}(1981)}]{EMRICH1981379}%
  \BibitemOpen
  \bibfield  {author} {\bibinfo {author} {\bibfnamefont {K.}~\bibnamefont
  {Emrich}},\ }\href {\doibase 10.1016/0375-9474(81)90179-2} {\bibfield
  {journal} {\bibinfo  {journal} {Nucl. Phys. A}\ }\textbf {\bibinfo {volume}
  {351}},\ \bibinfo {pages} {379 } (\bibinfo {year} {1981})}\BibitemShut
  {NoStop}%
\bibitem [{\citenamefont {{G{\"u}\ifmmode \mbox\c{c}\else \c{c}\fi{}l{\"u}}},
  \citenamefont {Wang},\ and\ \citenamefont {Guo}(2003)}]{PhysRevB.68.035304}%
  \BibitemOpen
  \bibfield  {author} {\bibinfo {author} {\bibfnamefont {A.~D.}\ \bibnamefont
  {{G{\"u}\ifmmode \mbox\c{c}\else \c{c}\fi{}l{\"u}}}}, \bibinfo {author}
  {\bibfnamefont {J.-S.}\ \bibnamefont {Wang}}, \ and\ \bibinfo {author}
  {\bibfnamefont {H.}~\bibnamefont {Guo}},\ }\href {\doibase
  10.1103/PhysRevB.68.035304} {\bibfield  {journal} {\bibinfo  {journal} {Phys.
  Rev. B}\ }\textbf {\bibinfo {volume} {68}},\ \bibinfo {pages} {035304}
  (\bibinfo {year} {2003})}\BibitemShut {NoStop}%
\bibitem [{\citenamefont {Pederiva}, \citenamefont {Umrigar},\ and\
  \citenamefont {Lipparini}(2000)}]{PhysRevB.62.8120}%
  \BibitemOpen
  \bibfield  {author} {\bibinfo {author} {\bibfnamefont {F.}~\bibnamefont
  {Pederiva}}, \bibinfo {author} {\bibfnamefont {C.~J.}\ \bibnamefont
  {Umrigar}}, \ and\ \bibinfo {author} {\bibfnamefont {E.}~\bibnamefont
  {Lipparini}},\ }\href@noop {} {\bibfield  {journal} {\bibinfo  {journal}
  {Phys. Rev. B}\ }\textbf {\bibinfo {volume} {62}},\ \bibinfo {pages} {8120}
  (\bibinfo {year} {2000})}\BibitemShut {NoStop}%
\bibitem [{\citenamefont {Bolton}(1996)}]{PhysRevB.54.4780}%
  \BibitemOpen
  \bibfield  {author} {\bibinfo {author} {\bibfnamefont {F.}~\bibnamefont
  {Bolton}},\ }\href {\doibase 10.1103/PhysRevB.54.4780} {\bibfield  {journal}
  {\bibinfo  {journal} {Phys. Rev. B}\ }\textbf {\bibinfo {volume} {54}},\
  \bibinfo {pages} {4780} (\bibinfo {year} {1996})}\BibitemShut {NoStop}%
\bibitem [{\citenamefont {Olsen}(2013)}]{olsen2013thesis}%
  \BibitemOpen
  \bibfield  {author} {\bibinfo {author} {\bibfnamefont {V.~K.~B.}\
  \bibnamefont {Olsen}},\ }\emph {\bibinfo {title} {Full {Configuration}
  {Interaction} {Simulation} of {Quantum} {Dots}}},\ \href
  {https://www.duo.uio.no/handle/10852/34217} {Master's thesis},\ \bibinfo
  {school} {University of Oslo} (\bibinfo {year} {2013})\BibitemShut {NoStop}%
\bibitem [{\citenamefont {Eto}(1997)}]{JJAP.36.3924}%
  \BibitemOpen
  \bibfield  {author} {\bibinfo {author} {\bibfnamefont {M.}~\bibnamefont
  {Eto}},\ }\href@noop {} {\bibfield  {journal} {\bibinfo  {journal} {Jpn. J.
  Appl. Phys.}\ }\textbf {\bibinfo {volume} {36}},\ \bibinfo {pages} {3924}
  (\bibinfo {year} {1997})}\BibitemShut {NoStop}%
\bibitem [{\citenamefont {Ezaki}, \citenamefont {Mori},\ and\ \citenamefont
  {Hamaguchi}(1997)}]{PhysRevB.56.6428}%
  \BibitemOpen
  \bibfield  {author} {\bibinfo {author} {\bibfnamefont {T.}~\bibnamefont
  {Ezaki}}, \bibinfo {author} {\bibfnamefont {N.}~\bibnamefont {Mori}}, \ and\
  \bibinfo {author} {\bibfnamefont {C.}~\bibnamefont {Hamaguchi}},\ }\href
  {\doibase 10.1103/PhysRevB.56.6428} {\bibfield  {journal} {\bibinfo
  {journal} {Phys. Rev. B}\ }\textbf {\bibinfo {volume} {56}},\ \bibinfo
  {pages} {6428} (\bibinfo {year} {1997})}\BibitemShut {NoStop}%
\bibitem [{\citenamefont {{Kvaal}}(2008)}]{2008arXiv0810.2644K}%
  \BibitemOpen
  \bibfield  {author} {\bibinfo {author} {\bibfnamefont {S.}~\bibnamefont
  {{Kvaal}}},\ }\href@noop {} {\bibfield  {journal} {\bibinfo  {journal} {ArXiv
  e-prints}\ } (\bibinfo {year} {2008})},\ \Eprint
  {http://arxiv.org/abs/0810.2644} {arXiv:0810.2644} \BibitemShut {NoStop}%
\bibitem [{\citenamefont {Rontani}\ \emph {et~al.}(2006)\citenamefont
  {Rontani}, \citenamefont {Cavazzoni}, \citenamefont {Bellucci},\ and\
  \citenamefont {Goldoni}}]{rontani:124102}%
  \BibitemOpen
  \bibfield  {author} {\bibinfo {author} {\bibfnamefont {M.}~\bibnamefont
  {Rontani}}, \bibinfo {author} {\bibfnamefont {C.}~\bibnamefont {Cavazzoni}},
  \bibinfo {author} {\bibfnamefont {D.}~\bibnamefont {Bellucci}}, \ and\
  \bibinfo {author} {\bibfnamefont {G.}~\bibnamefont {Goldoni}},\ }\href
  {\doibase 10.1063/1.2179418} {\bibfield  {journal} {\bibinfo  {journal} {J.
  Chem. Phys.}\ }\textbf {\bibinfo {volume} {124}},\ \bibinfo {pages} {124102}
  (\bibinfo {year} {2006})}\BibitemShut {NoStop}%
\bibitem [{\citenamefont {G{\l}azek}\ and\ \citenamefont
  {Wilson}(1993)}]{PhysRevD.48.5863}%
  \BibitemOpen
  \bibfield  {author} {\bibinfo {author} {\bibfnamefont {S.~D.}\ \bibnamefont
  {G{\l}azek}}\ and\ \bibinfo {author} {\bibfnamefont {K.~G.}\ \bibnamefont
  {Wilson}},\ }\href@noop {} {\bibfield  {journal} {\bibinfo  {journal} {Phys.
  Rev. D}\ }\textbf {\bibinfo {volume} {48}},\ \bibinfo {pages} {5863}
  (\bibinfo {year} {1993})}\BibitemShut {NoStop}%
\bibitem [{\citenamefont {Glazek}\ and\ \citenamefont
  {Wilson}(1994)}]{PhysRevD.49.4214}%
  \BibitemOpen
  \bibfield  {author} {\bibinfo {author} {\bibfnamefont {S.~D.}\ \bibnamefont
  {Glazek}}\ and\ \bibinfo {author} {\bibfnamefont {K.~G.}\ \bibnamefont
  {Wilson}},\ }\href@noop {} {\bibfield  {journal} {\bibinfo  {journal} {Phys.
  Rev. D}\ }\textbf {\bibinfo {volume} {49}},\ \bibinfo {pages} {4214}
  (\bibinfo {year} {1994})}\BibitemShut {NoStop}%
\bibitem [{\citenamefont {Anderson}\ \emph {et~al.}(2010)\citenamefont
  {Anderson}, \citenamefont {Bogner}, \citenamefont {Furnstahl},\ and\
  \citenamefont {Perry}}]{ScottSRG}%
  \BibitemOpen
  \bibfield  {author} {\bibinfo {author} {\bibfnamefont {E.~R.}\ \bibnamefont
  {Anderson}}, \bibinfo {author} {\bibfnamefont {S.~K.}\ \bibnamefont
  {Bogner}}, \bibinfo {author} {\bibfnamefont {R.~J.}\ \bibnamefont
  {Furnstahl}}, \ and\ \bibinfo {author} {\bibfnamefont {R.~J.}\ \bibnamefont
  {Perry}},\ }\href@noop {} {\bibfield  {journal} {\bibinfo  {journal} {Phys.
  Rev. C}\ }\textbf {\bibinfo {volume} {82}},\ \bibinfo {pages} {054001}
  (\bibinfo {year} {2010})}\BibitemShut {NoStop}%
\bibitem [{\citenamefont {Bogner}, \citenamefont {Furnstahl},\ and\
  \citenamefont {Perry}(2007)}]{PhysRevC.75.061001}%
  \BibitemOpen
  \bibfield  {author} {\bibinfo {author} {\bibfnamefont {S.~K.}\ \bibnamefont
  {Bogner}}, \bibinfo {author} {\bibfnamefont {R.~J.}\ \bibnamefont
  {Furnstahl}}, \ and\ \bibinfo {author} {\bibfnamefont {R.~J.}\ \bibnamefont
  {Perry}},\ }\href@noop {} {\bibfield  {journal} {\bibinfo  {journal} {Phys.
  Rev. C}\ }\textbf {\bibinfo {volume} {75}},\ \bibinfo {pages} {061001}
  (\bibinfo {year} {2007})}\BibitemShut {NoStop}%
\bibitem [{\citenamefont {{\AA}kerlund}\ \emph {et~al.}(2011)\citenamefont
  {{\AA}kerlund}, \citenamefont {Lindgren}, \citenamefont {Bergsten},
  \citenamefont {Grevholm}, \citenamefont {Lerner}, \citenamefont {Linscott},
  \citenamefont {Forssen},\ and\ \citenamefont {Platter}}]{SRGThreeDim}%
  \BibitemOpen
  \bibfield  {author} {\bibinfo {author} {\bibfnamefont {O.}~\bibnamefont
  {{\AA}kerlund}}, \bibinfo {author} {\bibfnamefont {E.}~\bibnamefont
  {Lindgren}}, \bibinfo {author} {\bibfnamefont {J.}~\bibnamefont {Bergsten}},
  \bibinfo {author} {\bibfnamefont {B.}~\bibnamefont {Grevholm}}, \bibinfo
  {author} {\bibfnamefont {P.}~\bibnamefont {Lerner}}, \bibinfo {author}
  {\bibfnamefont {R.}~\bibnamefont {Linscott}}, \bibinfo {author}
  {\bibfnamefont {C.}~\bibnamefont {Forssen}}, \ and\ \bibinfo {author}
  {\bibfnamefont {L.}~\bibnamefont {Platter}},\ }\href@noop {} {\bibfield
  {journal} {\bibinfo  {journal} {Eur. Phys. J. A}\ }\textbf {\bibinfo {volume}
  {47}},\ \bibinfo {pages} {122} (\bibinfo {year} {2011})}\BibitemShut
  {NoStop}%
\bibitem [{\citenamefont {Lohne}(2010)}]{lohne2010coupled}%
  \BibitemOpen
  \bibfield  {author} {\bibinfo {author} {\bibfnamefont {M.~P.}\ \bibnamefont
  {Lohne}},\ }\emph {\bibinfo {title} {Coupled-cluster studies of quantum
  dots}},\ \href {https://www.duo.uio.no/handle/10852/10966} {Master's
  thesis},\ \bibinfo  {school} {University of Oslo} (\bibinfo {year}
  {2010})\BibitemShut {NoStop}%
\bibitem [{{\relax DLMF}(2016)}]{NIST:DLMF}%
  \BibitemOpen
  {\relax DLMF},\ \href {http://dlmf.nist.gov/} {\enquote {\bibinfo {title}
  {{\it NIST Digital Library of Mathematical Functions}},}\ }\bibinfo
  {howpublished} {http://dlmf.nist.gov/, Release 1.0.14 of 2016-12-21}
  (\bibinfo {year} {2016}),\ \bibinfo {note} {f.~W.~J. Olver, A.~B. {Olde
  Daalhuis}, D.~W. Lozier, B.~I. Schneider, R.~F. Boisvert, C.~W. Clark, B.~R.
  Miller and B.~V. Saunders, eds.}\BibitemShut {Stop}%
\bibitem [{\citenamefont {Shavitt}\ and\ \citenamefont
  {Bartlett}(2009)}]{shavitt2009many}%
  \BibitemOpen
  \bibfield  {author} {\bibinfo {author} {\bibfnamefont {I.}~\bibnamefont
  {Shavitt}}\ and\ \bibinfo {author} {\bibfnamefont {R.~J.}\ \bibnamefont
  {Bartlett}},\ }\href@noop {} {\emph {\bibinfo {title} {{Many-Body Methods in
  Chemistry and Physics: MBPT and Coupled-Cluster Theory}}}},\ {Cambridge
  Molecular Science}\ (\bibinfo  {publisher} {Cambridge University Press},\
  \bibinfo {year} {2009})\BibitemShut {NoStop}%
\bibitem [{\citenamefont {Pulay}(1980)}]{PULAY1980393}%
  \BibitemOpen
  \bibfield  {author} {\bibinfo {author} {\bibfnamefont {P.}~\bibnamefont
  {Pulay}},\ }\href {\doibase 10.1016/0009-2614(80)80396-4} {\bibfield
  {journal} {\bibinfo  {journal} {Chem. Phys. Lett.}\ }\textbf {\bibinfo
  {volume} {73}},\ \bibinfo {pages} {393 } (\bibinfo {year}
  {1980})}\BibitemShut {NoStop}%
\bibitem [{\citenamefont {Pulay}(1982)}]{JCC:JCC540030413}%
  \BibitemOpen
  \bibfield  {author} {\bibinfo {author} {\bibfnamefont {P.}~\bibnamefont
  {Pulay}},\ }\href {\doibase 10.1002/jcc.540030413} {\bibfield  {journal}
  {\bibinfo  {journal} {J. Comput. Chem.}\ }\textbf {\bibinfo {volume} {3}},\
  \bibinfo {pages} {556} (\bibinfo {year} {1982})}\BibitemShut {NoStop}%
\bibitem [{\citenamefont {Broyden}(1965)}]{broyden1965class}%
  \BibitemOpen
  \bibfield  {author} {\bibinfo {author} {\bibfnamefont {C.~G.}\ \bibnamefont
  {Broyden}},\ }\href {\doibase 10.1090/S0025-5718-1965-0198670-6} {\bibfield
  {journal} {\bibinfo  {journal} {Math. Comput.}\ }\textbf {\bibinfo {volume}
  {19}},\ \bibinfo {pages} {577} (\bibinfo {year} {1965})}\BibitemShut
  {NoStop}%
\bibitem [{\citenamefont {Kehrein}(2006)}]{kehrein2006flow}%
  \BibitemOpen
  \bibfield  {author} {\bibinfo {author} {\bibfnamefont {S.}~\bibnamefont
  {Kehrein}},\ }\href@noop {} {\emph {\bibinfo {title} {{The Flow Equation
  Approach to Many-Particle Systems}}}},\ {Springer Tracts in Modern Physics}\
  (\bibinfo  {publisher} {Springer},\ \bibinfo {year} {2006})\BibitemShut
  {NoStop}%
\bibitem [{\citenamefont {Hjorth-Jensen}, \citenamefont {Lombardo},\ and\
  \citenamefont {van Kolck}(2017)}]{lnp936}%
  \BibitemOpen
  \bibfield  {author} {\bibinfo {author} {\bibfnamefont {M.}~\bibnamefont
  {Hjorth-Jensen}}, \bibinfo {author} {\bibfnamefont {M.~P.}\ \bibnamefont
  {Lombardo}}, \ and\ \bibinfo {author} {\bibfnamefont {U.}~\bibnamefont {van
  Kolck}},\ }\href {\doibase 10.1007/978-3-319-53336-0_1} {\emph {\bibinfo
  {title} {An Advanced Course in Computational Nuclear Physics: Bridging the
  Scales from Quarks to Neutron Stars}}},\ edited by\ \bibinfo {editor}
  {\bibfnamefont {M.}~\bibnamefont {Hjorth-Jensen}}, \bibinfo {editor}
  {\bibfnamefont {M.~P.}\ \bibnamefont {Lombardo}}, \ and\ \bibinfo {editor}
  {\bibfnamefont {U.}~\bibnamefont {van Kolck}},\ \bibinfo {series} {Lecture
  Notes in Physics}, Vol.\ \bibinfo {volume} {936}\ (\bibinfo  {publisher}
  {Springer International Publishing},\ \bibinfo {address} {Cham},\ \bibinfo
  {year} {2017})\BibitemShut {NoStop}%
\bibitem [{\citenamefont {Reimann}(2013)}]{reimann2013quantum}%
  \BibitemOpen
  \bibfield  {author} {\bibinfo {author} {\bibfnamefont {S.}~\bibnamefont
  {Reimann}},\ }\emph {\bibinfo {title} {Quantum-mechanical systems in traps
  and Similarity Renormalization Group theory}},\ \href
  {https://www.duo.uio.no/handle/10852/37161} {Master's thesis},\ \bibinfo
  {school} {University of Oslo} (\bibinfo {year} {2013})\BibitemShut {NoStop}%
\bibitem [{\citenamefont {Tsukiyama}, \citenamefont {Bogner},\ and\
  \citenamefont {Schwenk}(2012)}]{PhysRevC.85.061304}%
  \BibitemOpen
  \bibfield  {author} {\bibinfo {author} {\bibfnamefont {K.}~\bibnamefont
  {Tsukiyama}}, \bibinfo {author} {\bibfnamefont {S.~K.}\ \bibnamefont
  {Bogner}}, \ and\ \bibinfo {author} {\bibfnamefont {A.}~\bibnamefont
  {Schwenk}},\ }\href {\doibase 10.1103/PhysRevC.85.061304} {\bibfield
  {journal} {\bibinfo  {journal} {Phys. Rev. C}\ }\textbf {\bibinfo {volume}
  {85}},\ \bibinfo {pages} {061304} (\bibinfo {year} {2012})}\BibitemShut
  {NoStop}%
\bibitem [{\citenamefont {Tsukiyama}, \citenamefont {Bogner},\ and\
  \citenamefont {Schwenk}(2011)}]{PhysRevLett.106.222502}%
  \BibitemOpen
  \bibfield  {author} {\bibinfo {author} {\bibfnamefont {K.}~\bibnamefont
  {Tsukiyama}}, \bibinfo {author} {\bibfnamefont {S.~K.}\ \bibnamefont
  {Bogner}}, \ and\ \bibinfo {author} {\bibfnamefont {A.}~\bibnamefont
  {Schwenk}},\ }\href@noop {} {\bibfield  {journal} {\bibinfo  {journal} {Phys.
  Rev. Lett.}\ }\textbf {\bibinfo {volume} {106}},\ \bibinfo {pages} {222502}
  (\bibinfo {year} {2011})}\BibitemShut {NoStop}%
\bibitem [{\citenamefont {Roth}\ \emph {et~al.}(2012)\citenamefont {Roth},
  \citenamefont {Binder}, \citenamefont {Vobig}, \citenamefont {Calci},
  \citenamefont {Langhammer},\ and\ \citenamefont
  {Navr{\'a}til}}]{PhysRevLett.109.052501}%
  \BibitemOpen
  \bibfield  {author} {\bibinfo {author} {\bibfnamefont {R.}~\bibnamefont
  {Roth}}, \bibinfo {author} {\bibfnamefont {S.}~\bibnamefont {Binder}},
  \bibinfo {author} {\bibfnamefont {K.}~\bibnamefont {Vobig}}, \bibinfo
  {author} {\bibfnamefont {A.}~\bibnamefont {Calci}}, \bibinfo {author}
  {\bibfnamefont {J.}~\bibnamefont {Langhammer}}, \ and\ \bibinfo {author}
  {\bibfnamefont {P.}~\bibnamefont {Navr{\'a}til}},\ }\href@noop {} {\bibfield
  {journal} {\bibinfo  {journal} {Phys. Rev. Lett.}\ }\textbf {\bibinfo
  {volume} {109}},\ \bibinfo {pages} {052501} (\bibinfo {year}
  {2012})}\BibitemShut {NoStop}%
\bibitem [{\citenamefont {Morris}(2016)}]{morris2016thesis}%
  \BibitemOpen
  \bibfield  {author} {\bibinfo {author} {\bibfnamefont {T.~D.}\ \bibnamefont
  {Morris}},\ }\emph {\bibinfo {title} {Systematic improvements of
  \textit{ab-initio} in-medium similarity renormalization group
  calculations}},\ \href
  {https://publications.nscl.msu.edu/thesis/Morris_2016_387.pdf} {Ph.D.
  thesis},\ \bibinfo  {school} {Michigan State University} (\bibinfo {year}
  {2016})\BibitemShut {NoStop}%
\bibitem [{\citenamefont {Hugenholtz}(1957)}]{HUGENHOLTZ1957481}%
  \BibitemOpen
  \bibfield  {author} {\bibinfo {author} {\bibfnamefont {N.}~\bibnamefont
  {Hugenholtz}},\ }\href {\doibase 10.1016/S0031-8914(57)92950-6} {\bibfield
  {journal} {\bibinfo  {journal} {Physica}\ }\textbf {\bibinfo {volume} {23}},\
  \bibinfo {pages} {481 } (\bibinfo {year} {1957})}\BibitemShut {NoStop}%
\bibitem [{\citenamefont {Wick}(1950)}]{PhysRev.80.268}%
  \BibitemOpen
  \bibfield  {author} {\bibinfo {author} {\bibfnamefont {G.~C.}\ \bibnamefont
  {Wick}},\ }\href {\doibase 10.1103/PhysRev.80.268} {\bibfield  {journal}
  {\bibinfo  {journal} {Phys. Rev.}\ }\textbf {\bibinfo {volume} {80}},\
  \bibinfo {pages} {268} (\bibinfo {year} {1950})}\BibitemShut {NoStop}%
\bibitem [{\citenamefont {Bartlett}(1981)}]{ISI:A1981MN73700014}%
  \BibitemOpen
  \bibfield  {author} {\bibinfo {author} {\bibfnamefont {R.~J.}\ \bibnamefont
  {Bartlett}},\ }\href {\doibase 10.1146/annurev.pc.32.100181.002043}
  {\bibfield  {journal} {\bibinfo  {journal} {Annu. Rev. Phys. Chem.}\ }\textbf
  {\bibinfo {volume} {32}},\ \bibinfo {pages} {359} (\bibinfo {year}
  {1981})}\BibitemShut {NoStop}%
\bibitem [{\citenamefont {Shampine}\ and\ \citenamefont
  {Gordon}(1975)}]{shampine1975computer}%
  \BibitemOpen
  \bibfield  {author} {\bibinfo {author} {\bibfnamefont {L.}~\bibnamefont
  {Shampine}}\ and\ \bibinfo {author} {\bibfnamefont {M.}~\bibnamefont
  {Gordon}},\ }\href@noop {} {\emph {\bibinfo {title} {{Computer Solution of
  Ordinary Differential Equations: The Initial Value Problem}}}}\ (\bibinfo
  {publisher} {Freeman},\ \bibinfo {year} {1975})\BibitemShut {NoStop}%
\bibitem [{\citenamefont {Shampine}, \citenamefont {Gordon},\ and\
  \citenamefont {Burkardt}(2012)}]{odesolver}%
  \BibitemOpen
  \bibfield  {author} {\bibinfo {author} {\bibfnamefont {L.}~\bibnamefont
  {Shampine}}, \bibinfo {author} {\bibfnamefont {M.}~\bibnamefont {Gordon}}, \
  and\ \bibinfo {author} {\bibfnamefont {J.}~\bibnamefont {Burkardt}},\
  }\href@noop {} {\enquote {\bibinfo {title} {{ODE}: {Shampine} {and} {Gordon}
  {ODE} {Solver}},}\ }\bibinfo {howpublished}
  {\url{http://people.sc.fsu.edu/~jburkardt/c_src/ode/ode.html}} (\bibinfo
  {year} {2012})\BibitemShut {NoStop}%
\bibitem [{\citenamefont {Wegner}(2001)}]{Wegner200177}%
  \BibitemOpen
  \bibfield  {author} {\bibinfo {author} {\bibfnamefont {F.~J.}\ \bibnamefont
  {Wegner}},\ }\href {\doibase 10.1016/S0370-1573(00)00136-8} {\bibfield
  {journal} {\bibinfo  {journal} {Phys. Rep.}\ }\textbf {\bibinfo {volume}
  {348}},\ \bibinfo {pages} {77 } (\bibinfo {year} {2001})}\BibitemShut
  {NoStop}%
\bibitem [{\citenamefont {White}(2002)}]{White:cond-mat0201346}%
  \BibitemOpen
  \bibfield  {author} {\bibinfo {author} {\bibfnamefont {S.~R.}\ \bibnamefont
  {White}},\ }\href@noop {} {\bibfield  {journal} {\bibinfo  {journal} {J.
  Chem. Phys.}\ }\textbf {\bibinfo {volume} {117}},\ \bibinfo {pages} {7472}
  (\bibinfo {year} {2002})}\BibitemShut {NoStop}%
\bibitem [{\citenamefont {Neuscamman}, \citenamefont {Yanai},\ and\
  \citenamefont {Chan}(2010)}]{CTreview}%
  \BibitemOpen
  \bibfield  {author} {\bibinfo {author} {\bibfnamefont {E.}~\bibnamefont
  {Neuscamman}}, \bibinfo {author} {\bibfnamefont {T.}~\bibnamefont {Yanai}}, \
  and\ \bibinfo {author} {\bibfnamefont {G.~K.-L.}\ \bibnamefont {Chan}},\
  }\href {\doibase 10.1080/01442351003620540} {\bibfield  {journal} {\bibinfo
  {journal} {Int. Rev. Phys. Chem.}\ }\textbf {\bibinfo {volume} {29}},\
  \bibinfo {pages} {231} (\bibinfo {year} {2010})},\ \Eprint
  {http://arxiv.org/abs/https://doi.org/10.1080/01442351003620540}
  {https://doi.org/10.1080/01442351003620540} \BibitemShut {NoStop}%
\bibitem [{\citenamefont {Mazziotti}(2007{\natexlab{a}})}]{Mazziotti1}%
  \BibitemOpen
  \bibfield  {author} {\bibinfo {author} {\bibfnamefont {D.~A.}\ \bibnamefont
  {Mazziotti}},\ }\href {\doibase 10.1103/PhysRevA.75.022505} {\bibfield
  {journal} {\bibinfo  {journal} {Phys. Rev. A}\ }\textbf {\bibinfo {volume}
  {75}},\ \bibinfo {pages} {022505} (\bibinfo {year}
  {2007}{\natexlab{a}})}\BibitemShut {NoStop}%
\bibitem [{\citenamefont {Mazziotti}(2007{\natexlab{b}})}]{Mazziotti2}%
  \BibitemOpen
  \bibfield  {author} {\bibinfo {author} {\bibfnamefont {D.~A.}\ \bibnamefont
  {Mazziotti}},\ }\enquote {\bibinfo {title} {Anti-hermitian formulation of the
  contracted schr{\"o}dinger theory},}\ in\ \href {\doibase
  10.1002/9780470106600.ch12} {\emph {\bibinfo {booktitle}
  {Reduced-Density-Matrix Mechanics: With Application to Many-Electron Atoms
  and Molecules}}}\ (\bibinfo  {publisher} {John Wiley\& Sons, Inc.},\ \bibinfo
  {year} {2007})\ pp.\ \bibinfo {pages} {331--342}\BibitemShut {NoStop}%
\bibitem [{\citenamefont {Evangelista}(2014)}]{Evangelista}%
  \BibitemOpen
  \bibfield  {author} {\bibinfo {author} {\bibfnamefont {F.~A.}\ \bibnamefont
  {Evangelista}},\ }\href {\doibase 10.1063/1.4890660} {\bibfield  {journal}
  {\bibinfo  {journal} {J. Chem. Phys.}\ }\textbf {\bibinfo {volume} {141}},\
  \bibinfo {pages} {054109} (\bibinfo {year} {2014})}\BibitemShut {NoStop}%
\bibitem [{\citenamefont {Hergert}(2017)}]{HeikoReview}%
  \BibitemOpen
  \bibfield  {author} {\bibinfo {author} {\bibfnamefont {H.}~\bibnamefont
  {Hergert}},\ }\href {https://stacks.iop.org/1402-4896/92/i=2/a=023002}
  {\bibfield  {journal} {\bibinfo  {journal} {Phys. Scr.}\ }\textbf {\bibinfo
  {volume} {92}},\ \bibinfo {pages} {023002} (\bibinfo {year}
  {2017})}\BibitemShut {NoStop}%
\bibitem [{\citenamefont {Bartlett}\ and\ \citenamefont
  {Musia\l{}}(2007)}]{RevModPhys.79.291}%
  \BibitemOpen
  \bibfield  {author} {\bibinfo {author} {\bibfnamefont {R.~J.}\ \bibnamefont
  {Bartlett}}\ and\ \bibinfo {author} {\bibfnamefont {M.}~\bibnamefont
  {Musia\l{}}},\ }\href {\doibase 10.1103/RevModPhys.79.291} {\bibfield
  {journal} {\bibinfo  {journal} {Rev. Mod. Phys.}\ }\textbf {\bibinfo {volume}
  {79}},\ \bibinfo {pages} {291} (\bibinfo {year} {2007})}\BibitemShut
  {NoStop}%
\bibitem [{\citenamefont {{Hagen}}\ \emph {et~al.}(2014)\citenamefont
  {{Hagen}}, \citenamefont {{Papenbrock}}, \citenamefont {{Hjorth-Jensen}},\
  and\ \citenamefont {{Dean}}}]{2014RPPh...77i6302H}%
  \BibitemOpen
  \bibfield  {author} {\bibinfo {author} {\bibfnamefont {G.}~\bibnamefont
  {{Hagen}}}, \bibinfo {author} {\bibfnamefont {T.}~\bibnamefont
  {{Papenbrock}}}, \bibinfo {author} {\bibfnamefont {M.}~\bibnamefont
  {{Hjorth-Jensen}}}, \ and\ \bibinfo {author} {\bibfnamefont {D.~J.}\
  \bibnamefont {{Dean}}},\ }\href {\doibase 10.1088/0034-4885/77/9/096302}
  {\bibfield  {journal} {\bibinfo  {journal} {Rep. Prog. in Phys.}\ }\textbf
  {\bibinfo {volume} {77}},\ \bibinfo {eid} {096302} (\bibinfo {year}
  {2014})},\ \Eprint {http://arxiv.org/abs/1312.7872} {arXiv:1312.7872
  [nucl-th]} \BibitemShut {NoStop}%
\bibitem [{\citenamefont {Parzuchowski}, \citenamefont {Morris},\ and\
  \citenamefont {Bogner}(2017)}]{PhysRevC.95.044304}%
  \BibitemOpen
  \bibfield  {author} {\bibinfo {author} {\bibfnamefont {N.~M.}\ \bibnamefont
  {Parzuchowski}}, \bibinfo {author} {\bibfnamefont {T.~D.}\ \bibnamefont
  {Morris}}, \ and\ \bibinfo {author} {\bibfnamefont {S.~K.}\ \bibnamefont
  {Bogner}},\ }\href {\doibase 10.1103/PhysRevC.95.044304} {\bibfield
  {journal} {\bibinfo  {journal} {Phys. Rev. C}\ }\textbf {\bibinfo {volume}
  {95}},\ \bibinfo {pages} {044304} (\bibinfo {year} {2017})},\ \Eprint
  {http://arxiv.org/abs/1611.00661} {arXiv:1611.00661} \BibitemShut {NoStop}%
\bibitem [{\citenamefont {Anisimovas}\ and\ \citenamefont
  {Matulis}(1998)}]{0953-8984-10-3-013}%
  \BibitemOpen
  \bibfield  {author} {\bibinfo {author} {\bibfnamefont {E.}~\bibnamefont
  {Anisimovas}}\ and\ \bibinfo {author} {\bibfnamefont {A.}~\bibnamefont
  {Matulis}},\ }\href {\doibase 10.1088/0953-8984/10/3/013} {\bibfield
  {journal} {\bibinfo  {journal} {J. Phys. Condens. Mat.}\ }\textbf {\bibinfo
  {volume} {10}},\ \bibinfo {pages} {601} (\bibinfo {year} {1998})}\BibitemShut
  {NoStop}%
\bibitem [{\citenamefont {H{\o}gberget}(2013)}]{hoegberget2013thesis}%
  \BibitemOpen
  \bibfield  {author} {\bibinfo {author} {\bibfnamefont {J.}~\bibnamefont
  {H{\o}gberget}},\ }\emph {\bibinfo {title} {Quantum {Monte}-{Carlo} {Studies}
  of {Generalized} {Many}-body {Systems}}},\ \href
  {https://www.duo.uio.no/handle/10852/37167} {Master's thesis},\ \bibinfo
  {school} {University of Oslo} (\bibinfo {year} {2013})\BibitemShut {NoStop}%
\bibitem [{\citenamefont {Kutzelnigg}(1985)}]{Kutzelnigg1985}%
  \BibitemOpen
  \bibfield  {author} {\bibinfo {author} {\bibfnamefont {W.}~\bibnamefont
  {Kutzelnigg}},\ }\href {\doibase 10.1007/BF00527669} {\bibfield  {journal}
  {\bibinfo  {journal} {Theor. chim. acta}\ }\textbf {\bibinfo {volume} {68}},\
  \bibinfo {pages} {445} (\bibinfo {year} {1985})}\BibitemShut {NoStop}%
\bibitem [{\citenamefont {Klopper}\ and\ \citenamefont
  {Kutzelnigg}(1987)}]{KLOPPER198717}%
  \BibitemOpen
  \bibfield  {author} {\bibinfo {author} {\bibfnamefont {W.}~\bibnamefont
  {Klopper}}\ and\ \bibinfo {author} {\bibfnamefont {W.}~\bibnamefont
  {Kutzelnigg}},\ }\href {\doibase 10.1016/0009-2614(87)80005-2} {\bibfield
  {journal} {\bibinfo  {journal} {Chem. Phys. Lett.}\ }\textbf {\bibinfo
  {volume} {134}},\ \bibinfo {pages} {17 } (\bibinfo {year}
  {1987})}\BibitemShut {NoStop}%
\bibitem [{\citenamefont {Kong}, \citenamefont {Bischoff},\ and\ \citenamefont
  {Valeev}(2012)}]{explicitcorrelationreview}%
  \BibitemOpen
  \bibfield  {author} {\bibinfo {author} {\bibfnamefont {L.}~\bibnamefont
  {Kong}}, \bibinfo {author} {\bibfnamefont {F.~A.}\ \bibnamefont {Bischoff}},
  \ and\ \bibinfo {author} {\bibfnamefont {E.~F.}\ \bibnamefont {Valeev}},\
  }\href {\doibase 10.1021/cr200204r} {\bibfield  {journal} {\bibinfo
  {journal} {Chem. Rev.}\ }\textbf {\bibinfo {volume} {112}},\ \bibinfo {pages}
  {75} (\bibinfo {year} {2012})},\ \bibinfo {note} {pMID: 22176553}\BibitemShut
  {NoStop}%
\bibitem [{\citenamefont {Kvaal}(2009)}]{PhysRevB.80.045321}%
  \BibitemOpen
  \bibfield  {author} {\bibinfo {author} {\bibfnamefont {S.}~\bibnamefont
  {Kvaal}},\ }\href {\doibase 10.1103/PhysRevB.80.045321} {\bibfield  {journal}
  {\bibinfo  {journal} {Phys. Rev. B}\ }\textbf {\bibinfo {volume} {80}},\
  \bibinfo {pages} {045321} (\bibinfo {year} {2009})}\BibitemShut {NoStop}%
\bibitem [{\citenamefont {Kvaal}, \citenamefont {Hjorth-Jensen},\ and\
  \citenamefont {M{\o}ll~Nilsen}(2007)}]{Kvaal2007}%
  \BibitemOpen
  \bibfield  {author} {\bibinfo {author} {\bibfnamefont {S.}~\bibnamefont
  {Kvaal}}, \bibinfo {author} {\bibfnamefont {M.}~\bibnamefont
  {Hjorth-Jensen}}, \ and\ \bibinfo {author} {\bibfnamefont {H.}~\bibnamefont
  {M{\o}ll~Nilsen}},\ }\href {\doibase 10.1103/PhysRevB.76.085421} {\bibfield
  {journal} {\bibinfo  {journal} {Phys. Rev. B}\ }\textbf {\bibinfo {volume}
  {76}},\ \bibinfo {pages} {085421} (\bibinfo {year} {2007})}\BibitemShut
  {NoStop}%
\bibitem [{\citenamefont {Levenberg}(1944)}]{10.2307/43633451}%
  \BibitemOpen
  \bibfield  {author} {\bibinfo {author} {\bibfnamefont {K.}~\bibnamefont
  {Levenberg}},\ }\href {https://www.jstor.org/stable/43633451} {\bibfield
  {journal} {\bibinfo  {journal} {Q. Appl. Math.}\ }\textbf {\bibinfo {volume}
  {2}},\ \bibinfo {pages} {164} (\bibinfo {year} {1944})}\BibitemShut {NoStop}%
\bibitem [{\citenamefont {Marquardt}(1963)}]{doi:10.1137/0111030}%
  \BibitemOpen
  \bibfield  {author} {\bibinfo {author} {\bibfnamefont {D.~W.}\ \bibnamefont
  {Marquardt}},\ }\href {\doibase 10.1137/0111030} {\bibfield  {journal}
  {\bibinfo  {journal} {J. Soc. Ind. Appl. Math.}\ }\textbf {\bibinfo {volume}
  {11}},\ \bibinfo {pages} {431} (\bibinfo {year} {1963})}\BibitemShut
  {NoStop}%
\bibitem [{\citenamefont {Mor{\'e}}(1978)}]{More1978}%
  \BibitemOpen
  \bibfield  {author} {\bibinfo {author} {\bibfnamefont {J.~J.}\ \bibnamefont
  {Mor{\'e}}},\ }\enquote {\bibinfo {title} {The levenberg-marquardt algorithm:
  Implementation and theory},}\ in\ \href {\doibase 10.1007/BFb0067700} {\emph
  {\bibinfo {booktitle} {Numerical Analysis: Proceedings of the Biennial
  Conference Held at Dundee, June 28--July 1, 1977}}},\ \bibinfo {editor}
  {edited by\ \bibinfo {editor} {\bibfnamefont {G.~A.}\ \bibnamefont
  {Watson}}}\ (\bibinfo  {publisher} {Springer Berlin Heidelberg},\ \bibinfo
  {address} {Berlin, Heidelberg},\ \bibinfo {year} {1978})\ pp.\ \bibinfo
  {pages} {105--116}\BibitemShut {NoStop}%
\bibitem [{\citenamefont {Mor{\'e}}, \citenamefont {Garbow},\ and\
  \citenamefont {Hillstrom}(1980)}]{More:126569}%
  \BibitemOpen
  \bibfield  {author} {\bibinfo {author} {\bibfnamefont {J.~J.}\ \bibnamefont
  {Mor{\'e}}}, \bibinfo {author} {\bibfnamefont {B.~S.}\ \bibnamefont
  {Garbow}}, \ and\ \bibinfo {author} {\bibfnamefont {K.~E.}\ \bibnamefont
  {Hillstrom}},\ }\href {https://www.mcs.anl.gov/~more/ANL8074b.pdf} {\enquote
  {\bibinfo {title} {{User guide for MINPACK-1}},}\ }\bibinfo {type} {Tech.
  Rep.}\ \bibinfo {number} {ANL-80-74}\ (\bibinfo  {institution} {Argonne Nat.
  Lab.},\ \bibinfo {address} {Argonne, IL},\ \bibinfo {year}
  {1980})\BibitemShut {NoStop}%
\bibitem [{\citenamefont {Halkier}\ \emph {et~al.}(1999)\citenamefont
  {Halkier}, \citenamefont {Helgaker}, \citenamefont {J{\o}rgensen},
  \citenamefont {Klopper},\ and\ \citenamefont {Olsen}}]{HALKIER1999437}%
  \BibitemOpen
  \bibfield  {author} {\bibinfo {author} {\bibfnamefont {A.}~\bibnamefont
  {Halkier}}, \bibinfo {author} {\bibfnamefont {T.}~\bibnamefont {Helgaker}},
  \bibinfo {author} {\bibfnamefont {P.}~\bibnamefont {J{\o}rgensen}}, \bibinfo
  {author} {\bibfnamefont {W.}~\bibnamefont {Klopper}}, \ and\ \bibinfo
  {author} {\bibfnamefont {J.}~\bibnamefont {Olsen}},\ }\href {\doibase
  10.1016/S0009-2614(99)00179-7} {\bibfield  {journal} {\bibinfo  {journal}
  {Chem. Phys. Lett.}\ }\textbf {\bibinfo {volume} {302}},\ \bibinfo {pages}
  {437 } (\bibinfo {year} {1999})}\BibitemShut {NoStop}%
\bibitem [{\citenamefont {Morris}, \citenamefont {Parzuchowski},\ and\
  \citenamefont {Bogner}(2015)}]{PhysRevC.92.034331}%
  \BibitemOpen
  \bibfield  {author} {\bibinfo {author} {\bibfnamefont {T.~D.}\ \bibnamefont
  {Morris}}, \bibinfo {author} {\bibfnamefont {N.~M.}\ \bibnamefont
  {Parzuchowski}}, \ and\ \bibinfo {author} {\bibfnamefont {S.~K.}\
  \bibnamefont {Bogner}},\ }\href {\doibase 10.1103/PhysRevC.92.034331}
  {\bibfield  {journal} {\bibinfo  {journal} {Phys. Rev. C}\ }\textbf {\bibinfo
  {volume} {92}},\ \bibinfo {pages} {034331} (\bibinfo {year}
  {2015})}\BibitemShut {NoStop}%
\bibitem [{\citenamefont {Hergert}\ \emph {et~al.}(2013)\citenamefont
  {Hergert}, \citenamefont {Binder}, \citenamefont {Calci}, \citenamefont
  {Langhammer},\ and\ \citenamefont {Roth}}]{PhysRevLett.110.242501}%
  \BibitemOpen
  \bibfield  {author} {\bibinfo {author} {\bibfnamefont {H.}~\bibnamefont
  {Hergert}}, \bibinfo {author} {\bibfnamefont {S.}~\bibnamefont {Binder}},
  \bibinfo {author} {\bibfnamefont {A.}~\bibnamefont {Calci}}, \bibinfo
  {author} {\bibfnamefont {J.}~\bibnamefont {Langhammer}}, \ and\ \bibinfo
  {author} {\bibfnamefont {R.}~\bibnamefont {Roth}},\ }\href {\doibase
  10.1103/PhysRevLett.110.242501} {\bibfield  {journal} {\bibinfo  {journal}
  {Phys. Rev. Lett.}\ }\textbf {\bibinfo {volume} {110}},\ \bibinfo {pages}
  {242501} (\bibinfo {year} {2013})},\ \Eprint {http://arxiv.org/abs/1302.7294}
  {arXiv:1302.7294} \BibitemShut {NoStop}%
\bibitem [{\citenamefont {Hergert}\ \emph {et~al.}(2014)\citenamefont
  {Hergert}, \citenamefont {Bogner}, \citenamefont {Morris}, \citenamefont
  {Binder}, \citenamefont {Calci}, \citenamefont {Langhammer},\ and\
  \citenamefont {Roth}}]{PhysRevC.90.041302}%
  \BibitemOpen
  \bibfield  {author} {\bibinfo {author} {\bibfnamefont {H.}~\bibnamefont
  {Hergert}}, \bibinfo {author} {\bibfnamefont {S.~K.}\ \bibnamefont {Bogner}},
  \bibinfo {author} {\bibfnamefont {T.~D.}\ \bibnamefont {Morris}}, \bibinfo
  {author} {\bibfnamefont {S.}~\bibnamefont {Binder}}, \bibinfo {author}
  {\bibfnamefont {A.}~\bibnamefont {Calci}}, \bibinfo {author} {\bibfnamefont
  {J.}~\bibnamefont {Langhammer}}, \ and\ \bibinfo {author} {\bibfnamefont
  {R.}~\bibnamefont {Roth}},\ }\href {\doibase 10.1103/PhysRevC.90.041302}
  {\bibfield  {journal} {\bibinfo  {journal} {Phys. Rev. C}\ }\textbf {\bibinfo
  {volume} {90}},\ \bibinfo {pages} {041302} (\bibinfo {year} {2014})},\
  \Eprint {http://arxiv.org/abs/1408.6555} {arXiv:1408.6555} \BibitemShut
  {NoStop}%
\bibitem [{\citenamefont {Kutzelnigg}\ and\ \citenamefont
  {Mukherjee}(1997)}]{doi:10.1063/1.474405}%
  \BibitemOpen
  \bibfield  {author} {\bibinfo {author} {\bibfnamefont {W.}~\bibnamefont
  {Kutzelnigg}}\ and\ \bibinfo {author} {\bibfnamefont {D.}~\bibnamefont
  {Mukherjee}},\ }\href {\doibase 10.1063/1.474405} {\bibfield  {journal}
  {\bibinfo  {journal} {J. Chem. Phys.}\ }\textbf {\bibinfo {volume} {107}},\
  \bibinfo {pages} {432} (\bibinfo {year} {1997})}\BibitemShut {NoStop}%
\end{thebibliography}%
\end{document}